\documentclass{ws-procs9x6}
\usepackage{braket}
\usepackage{amsmath}	
\usepackage{color}

\begin{document}

\title{
QUANTUM ANNEALING:\\
FROM VIEWPOINTS OF STATISTICAL PHYSICS, CONDENSED MATTER PHYSICS,\\
AND COMPUTATIONAL PHYSICS
}

\author{SHU TANAKA}

\address{
Department of Chemistry, University of Tokyo,\\
7-3-1, Hongo, Bunkyo-ku, Tokyo, 113-0033, Japan\\
E-mail: shu-t@chem.s.u-tokyo.ac.jp
}

\author{RYO TAMURA}

\address{
Institute for Solid State Physics, University of Tokyo,\\
5-1-5, Kashiwanoha, Kashiwa-shi, Chiba, 277-8501, Japan\\
\vspace{1mm}
International Center for Young Scientists, National Institute for Materials Science,\\
1-2-1, Sengen, Tsukuba-shi, Ibaraki, 305-0047, Japan\\
E-mail: tamura.ryo@nims.go.jp
}

\begin{abstract}
In this paper, we review some features of quantum annealing and related topics from viewpoints of statistical physics, condensed matter physics, and computational physics.
We can obtain a better solution of optimization problems in many cases by using the quantum annealing.
Actually the efficiency of the quantum annealing has been demonstrated for problems based on statistical physics.
Then the quantum annealing has been expected to be an efficient and generic solver of optimization problems.
Since many implementation methods of the quantum annealing have been developed and will be proposed in the future, theoretical frameworks of wide area of science and experimental technologies will be evolved through studies of the quantum annealing.
\end{abstract}


\bodymatter

\section{Introduction}
\label{STsec:intro}

Optimization problems are present almost everywhere, for example, designing of integrated circuit, staff assignment, and selection of a mode of transportation.
To find the best solution of optimization problems is difficult in general.
Then, it is a significant issue to propose and to develop a method for obtaining the best solution (or a better solution) of optimization problems in information science.
In order to obtain the best solution, a couple of algorithms according to type of optimization problems have been formulated in information science and these methods have yielded practical applications.
Furthermore, since optimization problem is to find the state where a real-valued function takes the minimum value, it can be regarded as problem to obtain the ground state of the corresponding Hamiltonian.
Thus, if we can map optimization problem to well-defined Hamiltonian, we can use knowledge and methodologies of physics.
Actually, in computational physics, generic and powerful algorithms which can be adopted for wide application have been proposed.
One of famous methods is simulated annealing which was proposed by Kirkpatrick {\it et al.}\cite{STKirkpatrick-1983,STKirkpatrick-1984}.
In the simulated annealing, we introduce a temperature (thermal fluctuation) in the considered optimization problems.
We can obtain a better solution of the optimization problem by decreasing temperature gradually since thermal fluctuation effect facilitates transition between states.
It is guaranteed that we can obtain the best solution definitely if we decrease temperature slow enough\cite{STGeman-1984}.
Then, the simulated annealing has been used in many cases because of easy implementation and guaranty.

\begin{figure}
 \begin{center}
 \psfig{file=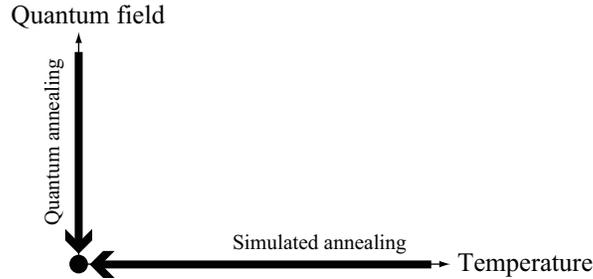,scale=1.0}
 \end{center}
 \caption{
 Schematic picture of the simulated annealing and the quantum annealing.
 Our purpose is to obtain the ground state at the point indicated by the solid circle.
 }
 \label{STfig:schematicSAQA}
\end{figure}

The quantum annealing was proposed as an alternative method of the simulated annealing\cite{STFinnila-1994,STKadowaki-1998a,STBrooke-1999,STFarhi-2001,STSantoro-2002,STDas-2005,STDas-2008,STOhzeki-2011}.
In the quantum annealing, we introduce a quantum field which is appropriate for the considered Hamiltonian.
For instance, if the considered optimization problem can be mapped onto the Ising model, the simplest form of the quantum fluctuation is transverse field.
In the quantum annealing, we gradually decrease quantum field (quantum fluctuation) instead of temperature (thermal fluctuation).
The efficiency of the quantum annealing has been demonstrated by a number of researchers, and it has been reported that a better solution can be obtained by the quantum annealing comparison with the simulated annealing in many cases.
Figure \ref{STfig:schematicSAQA} shows schematic picture of the simulated annealing and the quantum annealing.
In optimization problems, our target is to obtain the stable state at zero temperature and zero quantum field, which is indicated by the solid circle in Fig.~\ref{STfig:schematicSAQA}.

Recently, methods in which we decrease temperature and quantum field simultaneously have been proposed and as a result, we can obtain a better solution than the simulated annealing and the simple quantum annealing\cite{STKurihara-2009,STSato-2009,STTanaka-2011a}.
Moreover, as an another example of methods in which we use both thermal fluctuation and quantum fluctuation, novel quantum annealing method with the Jarzynski equality\cite{STJarzynski-1997a,STJarzynski-1997b} was also proposed\cite{STOhzeki-2010}, which is based on nonequilibrium statistical physics.

In this paper, we review the quantum annealing method which is the generic and powerful tool for obtaining the best solution of optimization problems from viewpoints of statistical physics, condensed matter physics, and computational physics.
The organization of this paper is as follows.
In Sec.~2, we review the Ising model which is a fundamental model of magnetic systems.
The realization method of the Ising model by nuclear magnetic resonance is also explained.
In Sec.~3, we show a couple of implementation methods of the quantum annealing.
In Sec.~4, we explain two optimization problems -- traveling salesman problem and clustering problem.
The quantum annealing based on the Monte Carlo method for the traveling salesman problem is also demonstrated.
In Sec.~5, we review related topics of the quantum annealing -- Kibble-Zurek mechanism of the Ising spin chain and order by disorder in frustrated systems. 
In Sec.~6, we summarize this paper briefly and give some future perspectives of the quantum annealing.
\section{Ising Model}
\label{STsec:Ising}

In this section we introduce the Ising model which is a fundamental model in statistical physics.
A century ago, the Ising model was proposed to explain cooperative nature in strongly correlated magnetic systems from a microscopic viewpoint\cite{STIsing-1925}.
The Hamiltonian of the Ising model is given by
\begin{eqnarray}
 {\cal H}_{\rm Ising} = - \sum_{i,j} J_{ij} \sigma_i^z \sigma_j^z - \sum_{i=1}^N h_i \sigma_i^z,
  \qquad
  \sigma_i^z = \pm 1,
\end{eqnarray}
where the summation of the first term runs over all interactions on the defined graph and $N$ represents the number of spins.
If the sign of $J_{ij}$ is positive/negative, the interaction is called ferromagnetic/antiferromagnetic interaction.
Spins which are connected by ferromagnetic/antiferromagnetic interaction tend to be the same/opposite direction.
The second term of the Hamiltonian denotes the site-dependent longitudinal magnetic fields.
Although the Ising model is quite simple, this model exhibits inherent rich properties {\it e.g.} phase transition and dynamical behavior such as melting process and slow relaxation.
For instance, the ferromagnetic Ising model with homogeneous interaction ($J_{ij}=J$ for $\forall i,j$) and no external magnetic fields ($h_i = 0$ for $\forall i$) on square lattice exhibits the second-order phase transition, whereas no phase transition occurs in the Ising model on one-dimensional lattice.
Onsager first succeeded to obtain explicitly free energy of the Ising model without external magnetic field on square lattice\cite{STOnsager-1944}.
After that, a couple of calculation methods were proposed.
Furthermore, these calculation methods have been improved day by day, and the new techniques which were developed in these methods have been applied for other more complicated problems.
Since the Ising model is quite simple, we can easily generalize the Ising model in diverse ways such as the Blume-Capel model\cite{STBlume-1966,STCapel-1966}, the clock model\cite{STTobochnik-1982,STChalla-1986}, and the Potts model\cite{STPotts-1952,STWu-1982}.
By analyzing these models, relation between nature of phase transition and the symmetry which breaks at the transition point has been investigated.
Then, it is not too much to say that the Ising model has opened up a new horizon for statistical physics.

The Ising model can be adopted for not only magnetic systems but also systems in wide area of science such as information science.
Optimization problem is one of important topics in information science.
As we mention in Sec.~\ref{STsec:optimization_problems}, optimization problem can be mapped onto the Ising model and its generalized models in many cases.
Then some methods which were developed in statistical physics often have been used for optimization problem.
In Sec.~\ref{STsec:magneticsystems_ising}, we show a couple of magnetic systems which can be well represented by the Ising model.
In Sec.~\ref{STsec:nmr_ising}, we review how to create the Ising model by Nuclear Magnetic Resonance (NMR) technique as an example of experimental realization of the Ising model.

\subsection{Magnetic Systems}
\label{STsec:magneticsystems_ising}

In many cases, the Hamiltonian of magnetic systems without external magnetic field is given by
\begin{eqnarray}
\nonumber
 \hat{{\cal H}} &&= - \sum_{i,j} J_{ij} \hat{\sigma}_i \cdot \hat{\sigma}_j\\
 &&= - \sum_{i,j} J_{ij} 
  \left(
   \hat{\sigma}_i^x \cdot \hat{\sigma}_j^x +
   \hat{\sigma}_i^y \cdot \hat{\sigma}_j^y +
   \hat{\sigma}_i^z \cdot \hat{\sigma}_j^z 
  \right),
\end{eqnarray}
where $\hat{\sigma}_i^\alpha$ denotes the $\alpha$-component of the Pauli matrix at the $i$-th site.
The form of this interaction is called Heisenberg interaction.
The definitions of Pauli matrices are 
\begin{eqnarray}
 \hat{\sigma}^x :=
  \left(
   \begin{array}{cc}
    0 & 1\\
    1 & 0
   \end{array}
  \right),
  \qquad
 \hat{\sigma}^y :=
  \left(
   \begin{array}{cc}
   0 & -i\\
   i & 0
   \end{array}
  \right),
  \qquad
 \hat{\sigma}^z :=
  \left(
   \begin{array}{cc}
   1 & 0\\
   0 & -1
   \end{array}
  \right),
\end{eqnarray}
where the bases are defined by
\begin{eqnarray}
 \ket{\uparrow}:=
  \left(
   \begin{array}{c}
    1\\
    0
   \end{array}
  \right),
  \qquad
 \ket{\downarrow}:=
  \left(
   \begin{array}{c}
    0\\
    1
   \end{array}
  \right).
  \qquad
\end{eqnarray}
In this case, magnetic interactions are isotropic.
However, they become anisotropic depending on the surrounded ions in real magnetic materials.
In general, the Hamiltonian of magnetic systems should be replaced by
\begin{eqnarray}
 \hat{{\cal H}} = - \sum_{i,j} J_{ij}
  \left(
   c_x \hat{\sigma}_i^x \hat{\sigma}_j^x +
   c_y \hat{\sigma}_i^y \hat{\sigma}_j^y +
   c_z \hat{\sigma}_i^z \hat{\sigma}_j^z 
  \right).
\end{eqnarray}
When $|c_x|, |c_y| > |c_z|$, the $xy$-plane is easy-plane and the Hamiltonian becomes XY-like Hamiltonian.
On the contrary, when $|c_z| > |c_x|, |c_y|$, the $z$-axis is easy-axis and the Hamiltonian becomes Ising-like Hamiltonian.
Such anisotropy comes from crystal structure, spin-orbit coupling, and dipole-dipole coupling.
Moreover, even if there is almost no anisotropy in magnetic interactions, magnetic systems can be regarded as the Ising model when the number of electrons in the magnetic ion is odd and the total spin is half-integer.
In this case, doubly degenerated states exist because of the Kramers theorem.
These states are called the Kramers doublet.
When the energy difference between the ground states and the first-excited states $\Delta E$ is large enough, these doubly-degenerated ground states can be well represented by the $S=1/2$ Ising spins.
Table~\ref{STtable:Isingmaterials} shows examples of the magnetic materials which can be well represented by the Ising model on one-dimensional chain, two-dimensional square lattice, and three-dimensional cubic lattice.

\begin{sidewaystable}
\tbl{
Examples of magnetic materials which can be represented by the Ising model on chain (one-dimension), square lattice (two-dimension), and cubic lattice (three-dimension).
}
{
\begin{tabular}{c|ccccc}
\toprule
\\[-8pt] Material & Spatial dimension & Total spin & Type of interaction & $J/k_\text{B}$ & References  \\[-8pt]\\ \hline\hline
\\[-8pt] K$_3$Fe(CN)$_6$ & One (chain) & $\frac{1}{2}$ & Antiferromagnetic & $-0.23$ K & \refcite{STOhtsuka-1961,STRayl-1968,STOno-1970} \\[-8pt]\\ \hline
\\[-8pt] CsCoCl$_3$ & One (chain) & $\frac{1}{2}$ & Antiferromagnetic & $-100$ K & \refcite{STAchiwa-1969, STMekata-1978} \\[-8pt]\\ \hline
\\[-8pt] Dy(C$_2$H$_5$SO$_4$)$_2$ $\cdot$ 9 H$_2$O & One (chain) & $\frac{1}{2}$ & Ferromagnetic & $0.2$ K & \refcite{STCooke-1959,STCooke-1968a,STCooke-1968b} \\[-8pt]\\ \hline
\\[-8pt] CoCl$_2$ $\cdot$ 2NC$_5$H$_5$ & One (chain) & $\frac{1}{2}$ & Ferromagnetic & $9.5$ K & \refcite{STTakeda-1970,STTakeda-1971a} \\[-8pt]\\ \hline
\hline
\\[-8pt] CoCs$_3$Br$_5$ & Two (square) & $\frac{1}{2}$ & Antiferromagnetic & $-0.23$ K & \refcite{STFiggis-1964,STWielinga-1967,STMess-1967} \\[-8pt]\\ \hline
\\[-8pt] Co(HCOO)$_2$ $\cdot$ 2 H$_2$O & Two (square) & $\frac{1}{2}$ & Antiferromagnetic & $-4.3$ K & \refcite{STHoy-1965,STMatsuura-1970,STPierce-1971,STTakeda-1971b} \\[-8pt]\\ \hline
\\[-8pt] Rb$_2$CoF$_4$ & Two (square) & $\frac{1}{2}$ & Antiferromagnetic & $-91$ K & \refcite{STStryiewski-1977,STBreed-1969} \\[-8pt]\\ \hline
\\[-8pt] FeCl$_2$ & Two (square) & $1$ & Ferromagnetic & $3.4$ K & \refcite{STOno-1964,STBirgeneau-1972} \\[-8pt]\\ \hline
\hline
\\[-8pt] DyPO$_4$ & Three (cubic) & $\frac{1}{2}$ & Antiferromagnetic & $-2.5$ K & \refcite{STWright-1971,STRado-1969,STScharenberg-1971,STFuess-1971} \\[-8pt]\\ \hline
\\[-8pt] Dy$_3$Al$_5$O$_{12}$ & Three (cubic)& $\frac{1}{2}$ & Antiferromagnetic & $-1.85$ K & \refcite{STBall-1963,STNorvell-1969a,STNorvell-1969b} \\[-8pt]\\ \hline
\\[-8pt] CoRb$_3$Cl$_5$ & Three (cubic)& $\frac{1}{2}$ & Antiferromagnetic & $-0.511$ K & \refcite{STBaker-1963,STSykes-1972} \\[-8pt]\\ \hline
\\[-8pt] FeF$_2$ & Three (cubic)& $2$ & Antiferromagnetic & $-2.69$ K & \refcite{STStout-1955,STDomb-1964,STWertheim-1967,STShapira-1970} \\[-8pt]\\
\hline\hline
\end{tabular}}
\label{STtable:Isingmaterials}
\end{sidewaystable}

\subsection{Nuclear Magnetic Resonance}
\label{STsec:nmr_ising}

In condensed matter physics, Nuclear Magnetic Resonance (NMR) has been used for decision of the structure of organic compounds and for analysis of the state in materials by using resonance induced by electromagnetic wave.
The NMR can {\it create} the Ising model with transverse fields, which is expected to become an element of quantum information processing.
In this processing, we use molecules where the coherence times are long compared with typical gate operations.
Actually a couple of molecules which have nuclear spins were used for demonstration of quantum computing\cite{STNielsen-2000,STNakahara-2008,STCory-1997,STCory-1998,STGershenfeld-1997,STChuang-1998,STJones-1998,STKnill-1998,STLaflamme-1998,STJones-1999,STPrice-1999,STVandersypen-1999,STVandersypen-2000,STVandersypen-2001,STNakahara-2004,STKondo-2007}.
In this section we explain how to create the Ising model by NMR.

The setup of the NMR spectrometer as a tool of quantum computing is as follows.
We first put molecules which contain nuclear spins under the strong magnetic field $B_0$.
Next we apply radio frequency $\omega^{\rm (rf)}$ magnetic field which is perpendicular to the strong magnetic field $B_0$.
For simplicity, we here consider a molecule which contains two spins.
We also assume that the considered molecule can be well described by the Heisenberg Hamiltonian.
Then the Hamiltonian of this system is given by
\begin{eqnarray}
 \hat{{\cal H}} = \hat{{\cal H}}_{\rm mol} + \hat{{\cal H}}_1^{\rm (rf)} + \hat{{\cal H}}_2^{\rm (rf)},
\end{eqnarray}
where $\hat{\cal H}_{\rm mol}$, $\hat{\cal H}_1^{\rm (rf)}$, and $\hat{\cal H}_2^{\rm (rf)}$ are defined by
\begin{eqnarray}
 &&\hat{\cal H}_{\rm mol} := -h_1 \hat{\sigma}_1^z - h_2 \hat{\sigma}_2^z 
  - J (\hat{\sigma}_1^x \cdot \hat{\sigma}_2^x + \hat{\sigma}_1^y \cdot \hat{\sigma}_2^y + \hat{\sigma}_1^z \cdot \hat{\sigma}_2^z),\\
 &&\hat{\cal H}_{1}^{\rm (rf)} := -\Gamma_1 \cos (\omega^{\rm (rf)} t - \phi_1) (\hat{\sigma}_1^x + \gamma' \hat{\sigma}_2^x),\\
 &&\hat{\cal H}_{2}^{\rm (rf)} := -\Gamma_2 \cos (\omega^{\rm (rf)} t - \phi_2) (\gamma'^{-1}\hat{\sigma}_1^x + \hat{\sigma}_2^x),
\end{eqnarray}
respectively.
We take the natural unit in which $\hbar = 1$.
The values of $\phi_1$ and $\phi_2$ are the phases at the time $t=0$ of the first spin and that of the second spin, respectively.
The quantities of $h_i$ are defined by $h_i:=\gamma_i B_0$, where $\gamma_i$ denotes the gyromagnetic ratio of the $i$-th spin ($i=1,2$).
The values of $h_1$ and $h_2$ represent energy differences between $\ket{\uparrow}$ and $\ket{\downarrow}$ of the first spin and the second spin, respectively.
The coefficients $\Gamma_1$ and $\Gamma_2$ in $\hat{\cal H}_1^{\rm (rf)}$ and $\hat{\cal H}_2^{\rm (rf)}$ are the effective amplitudes of the ac magnetic field, whose definitions are $\Gamma_i := \gamma_i B_{\rm ac}$, where $B_{\rm ac}$ is amplitude of the ac magnetic field.
The value of $\gamma'$ is defined by the ratio of the gyromagnetic ratios $\gamma':= \gamma_2/\gamma_1$.

We define the following unitary transformation:
\begin{eqnarray}
 \hat{U}^{\rm (R)} := {\rm e}^{-ih_1\hat{\sigma}_1^z t} \cdot {\rm e}^{-ih_2\hat{\sigma}_2^z t}.
\end{eqnarray}
We can change from the laboratory frame to a frame rotating with $h_i$ around the $z$-axis by using the above unitary transformation.
The dynamics of a density matrix can be calculated by
\begin{eqnarray}
 \label{STeq:LNeq_normal}
 i \frac{{\rm d}\hat{\rho}}{{\rm d}t} = [\hat{\cal H},\hat{\rho}].
\end{eqnarray}
The density matrix on the rotating frame is given by
\begin{eqnarray}
 \hat{\rho}^{\rm (R)} := \hat{U}^{\rm (R)} \hat{\rho} \hat{U}^{{\rm (R)}\dagger}.
\end{eqnarray}
To be the same form as Eq.~(\ref{STeq:LNeq_normal}) on the rotating frame, the Hamiltonian on the rotating frame should be
\begin{eqnarray}
 \hat{\cal H}^{\rm (R)} = \hat{U}^{\rm (R)} \hat{\cal H} \hat{U}^{{\rm (R)}\dagger}
  - i \hat{U}^{\rm (R)} \frac{{\rm d}\hat{U}^{{\rm (R)}\dagger}}{{\rm d}t}.
\end{eqnarray}
Here we decompose the Hamiltonian on the rotating frame as
\begin{eqnarray}
 \hat{\cal H}^{\rm (R)} = \hat{\cal H}_{\rm mol}^{\rm (R)} + \hat{\cal H}_1^{\rm (R)(rf)} + \hat{\cal H}_2^{\rm (R)(rf)},
\end{eqnarray}
where the three terms are defined by
\begin{eqnarray}
 &&\hat{\cal H}_{\rm mol}^{\rm (R)} := \hat{U}^{\rm (R)} \hat{\cal H}_{\rm mol} \hat{U}^{{\rm (R)}\dagger}
  - i \hat{U}^{\rm (R)} \frac{{\rm d}\hat{U}^{{\rm (R)}\dagger}}{{\rm d}t},\\
 &&\hat{\cal H}_1^{\rm (R)(rf)} := \hat{U}^{\rm (R)} \hat{\cal H}_1^{\rm (rf)} \hat{U}^{{\rm (R)}\dagger},\\
 &&\hat{\cal H}_2^{\rm (R)(rf)} := \hat{U}^{\rm (R)} \hat{\cal H}_2^{\rm (rf)} \hat{U}^{{\rm (R)}\dagger}.
\end{eqnarray}
The intramolecular magnetic interaction Hamiltonian on the rotating frame $\hat{\cal H}_{\rm mol}^{\rm (R)}$ can be calculated as
\begin{eqnarray}
 \hat{\cal H}_{\rm mol}^{\rm (R)} = 
  J 
  \left(
   \begin{array}{cccc}
   0 & 0& 0& 0\\
   0 & 0& {\rm e}^{i (h_2 - h_1) t}& 0\\
   0 & {\rm e}^{-i (h_2 - h_1) t} & 0& 0\\
   0 & 0& 0& 0
   \end{array}
  \right)
  - J \hat{\sigma}_1^z \hat{\sigma}_2^z 
  \simeq - J \hat{\sigma}_1^z \hat{\sigma}_2^z.
\end{eqnarray}
The approximation is valid when $|h_2 - h_1|\tau \gg 1$, where $\tau$ is a characteristic time scale since the exponential terms are averaged to vanish.
The radio frequency magnetic field Hamiltonian on the rotating frame $\hat{\cal H}_1^{\rm (R)(rf)}$ under the resonance condition $\omega^{\rm (rf)}=h_i$ can be calculated as
\begin{eqnarray}
 \nonumber
 \hat{\cal H}_1^{\rm (R)(rf)} = 
   -\Gamma_1
   \left[
    \left(
     \begin{array}{cccc}
      0& 0&{\rm e}^{-i\phi_1} &0 \\
      0& 0& 0&{\rm e}^{-i\phi_1} \\
      {\rm e}^{i\phi_1}& 0& 0& 0\\
      0 &{\rm e}^{i\phi_1} & 0& 0\\
     \end{array}
    \right) 
    + \gamma' 
    \left(
     \begin{array}{cccc}
      0& a_{--}& 0& 0\\
      a_{++} &0 &0 &0 \\
      0&0 &0 & a_{--}\\
      0&0 & a_{++} & 0
     \end{array}
    \right)
   \right],
\end{eqnarray}
where $a_{--}:= {\rm e}^{-i(h_2-h_1)t+\phi_1}+{\rm e}^{-i(h_1+h_2)t-\phi_1}$ and $a_{++}:={\rm e}^{i(h_2-h_1)t+\phi_1}+{\rm e}^{i(h_1+h_2)t-\phi_1}$.
The second term of $\hat{\cal H}_1^{\rm (R)(rf)}$ vanishes when $|h_1+h_2|\tau, |h_2-h_1|\tau \gg 1$.
Then under these conditions, the Hamiltonian becomes
\begin{eqnarray}
 \hat{\cal H}_1^{\rm (R)(rf)} = - \Gamma_1 
  ( \cos\phi_1 \hat{\sigma}_1^x + \sin\phi_1 \hat{\sigma}_1^y).
\end{eqnarray}
In the same way, the Hamiltonian $\hat{\cal H}_2^{\rm (R)(rf)}$ can be calculated as
\begin{eqnarray}
 \hat{\cal H}_2^{\rm (R)(rf)} = - \Gamma_2 
  ( \cos\phi_2 \hat{\sigma}_2^x + \sin\phi_2 \hat{\sigma}_2^y).
\end{eqnarray}
By taking the rotation operators on the individual sites, we can rewrite the Hamiltonians $\hat{\cal H}_1^{\rm (R)(rf)}$ and $\hat{\cal H}_2^{\rm (R)(rf)}$ by only the $x$-component of the Pauli matrix:
\begin{eqnarray}
 {\rm e}^{i \phi_1 \hat{\sigma}_1^z} \hat{\cal H}_1^{\rm (R)(rf)} {\rm e}^{-i \phi_1 \hat{\sigma}_1^z} &= -\Gamma_1 \hat{\sigma}_1^x,\\
{\rm e}^{i \phi_2 \hat{\sigma}_2^z} \hat{\cal H}_2^{\rm (R)(rf)} {\rm e}^{-i \phi_2 \hat{\sigma}_2^z} &= -\Gamma_2 \hat{\sigma}_2^x. 
\end{eqnarray}
Then, the total Hamiltonian can be represented by the Ising model with site-dependent transverse fields:
\begin{eqnarray}
 \hat{\cal H}^{\rm (R)} = -J \hat{\sigma}_1^z \hat{\sigma}_2^z 
  -\Gamma_1 \hat{\sigma}_1^x - \Gamma_2 \hat{\sigma}_2^x.
\end{eqnarray}
It should be noted that the above procedure is not restricted for two spin system.
Then, the NMR technique can be create the Ising model with site-dependent transverse fields in general.
\section{Implementation Methods of Quantum Annealing}
\label{STsec:implementation}

As stated in Sec.~\ref{STsec:intro}, the quantum annealing method is expected to be a powerful tool to obtain the best solution of optimization problems in a generic way.
The quantum annealing methods can be categorized according to how to treat time-development.
One is a stochastic method such as the Monte Carlo method which will be shown in Sec.~\ref{STsubsec:MC}.
Other is a deterministic method such as mean-field type method and real-time dynamics.
We will explain the mean-field type method and the method based on real-time dynamics in Secs.~\ref{STsubsec:MF} and \ref{STsubsec:RT}.
Although in the Monte Carlo method and the mean-field type method, we introduce time-development in an artificial way, the merit of these methods is to be able to treat large-scale systems.
The methods based on the Schr\"odinger equation can follow up real-time dynamics which occurs in real experimental systems.
However, these methods can be used for very small systems and/or limited lattice geometries because of limited computer resources and characters of algorithms.
Each method has strengths and limitations based on its individuality.
Then when we use the quantum annealing, we have to choose implementation methods according to what we want to know.
In this section, we explain three types of theoretical methods for the quantum annealing and some experimental results which relate to the quantum annealing.

\subsection{Monte Carlo Method}
\label{STsubsec:MC}

In this section we review the Monte Carlo method as an implementation method of the quantum annealing.
In physics, the Monte Carlo method is widely adopted for analysis of equilibrium properties of strongly correlated systems such as spin systems, electric systems, and bosonic systems.
Originally the Monte Carlo method is used in order to calculate integrated value of given function.
The simplest example is ``calculation of $\pi$''.
Suppose we consider a square in which $-1 \le x,y \le 1$ and a circle whose radius is unity and center is $(x,y)=(0,0)$.
We generate pair of uniform random numbers ($-1 \le x_i,y_i \le 1$) many times and calculate the following quantity:
\begin{eqnarray}
 \frac{{\rm number\,\, of\,\, steps\,\, when\,\,}\sqrt{x_i^2+y_i^2}\le 1{\rm \,\, is\,\, satisfied}}{{\rm number\,\, of\,\, steps}}.
\end{eqnarray}
Hereafter we refer to the denominator as Monte Carlo step.
The quantity should converge to $\pi/4$ in the limit of infinite Monte Carlo step.
This is a pedagogical example of the Monte Carlo method.
We first explain how to implement and theoretical background of the Monte Carlo method which is used in physics.

In equilibrium statistical physics, we would like to know the equilibrium value at given temperature $T$.
The equilibrium value of the physical quantity which is represented by the operator $O$ is defined as
\begin{eqnarray}
 \label{STeq:eqvalue_O_def}
 \langle O \rangle_T ^{\rm (eq)} :=
  \frac{
  {\rm Tr}\, O {\rm e}^{-\beta{\cal H}}
  }{
  {\rm Tr}\, {\rm e}^{-\beta{\cal H}}
  },
\end{eqnarray}
where ${\rm Tr}$ means the trace of matrix and $\beta$ denotes the inverse temperature $\beta = (k_{\rm B}T)^{-1}$.
Hereafter we set the Boltzmann constant $k_{\rm B}$ to be unity.
For small systems, we can obtain the equilibrium value by taking sum analytically, on the contrary, it is difficult to obtain the equilibrium value for large systems except few solvable models.
Then in order to evaluate equilibrium value of the physical quantity, we often use the Monte Carlo method.

We consider the Ising model given by
\begin{eqnarray}
 {\cal H}_{\rm Ising} = - \sum_{\langle i,j \rangle} J_{ij} \sigma_i^z \sigma_j^z
  - \sum_{i=1}^N h_i \sigma_i^z, \qquad \sigma_i^z=\pm 1.
\end{eqnarray}
The Ising model without transverse field can be expressed as a diagonal matrix by using ``trivial'' bit representation $\ket{\uparrow}$ and $\ket{\downarrow}$ which were introduced in Sec.~\ref{STsec:Ising}.
Then, in this case, we can easily calculate the eigenenergy once the eigenstate is specified.

We can use the Monte Carlo method for obtaining the equilibrium value defined by Eq.~(\ref{STeq:eqvalue_O_def}) as well as the calculation of $\pi$:
\begin{eqnarray}
 \label{STeq:eqvalue_O_simple}
 \frac{
  \sum_\Sigma O(\Sigma) {\rm e}^{-\beta E(\Sigma)}
  }{
  \sum_\Sigma {\rm e}^{-\beta E(\Sigma)}
  }
  \to
  \langle O \rangle_T ^{\rm (eq)},
\end{eqnarray}
where $O(\Sigma)$ and $E (\Sigma)$ denote the physical value of $O$ and the eigenenergy of the eigenstate $\Sigma$, respectively.
Here the eigenstate $\Sigma$ is generated by uniform random number and $\sum_\Sigma 1$ is equal to Monte Carlo step.
In the limit of infinite Monte Carlo step, LHS of Eq.~(\ref{STeq:eqvalue_O_simple}) should be converge to the equilibrium value.
Equilibrium statistical physics says that the probability distribution at equilibrium state can be described by the Boltzmann distribution which is proportional to ${\rm e}^{-\beta E(\Sigma)}$.
In this case, since we know the form of the probability distribution, it is better to use the distribution function to generate a state according to the Boltzmann distribution instead of uniform random number.
This scheme is called importance sampling.
When we use the importance sampling, we can obtain the equilibrium value as follows:
\begin{eqnarray}
 \frac{
  \sum_\Sigma O(\Sigma)
  }{
  \sum_\Sigma 1
  }
  \to 
  \langle O \rangle_T ^{\rm (eq)}.
\end{eqnarray}
In order to generate a state according to the Boltzmann distribution, we use the Markov chain Monte Carlo method.
Let $P(\Sigma_a,t)$ be the probability of the $a$-th state at time $t$.
In this method, time-evolution of probability distribution is given by the master equation:
\begin{eqnarray}
 \nonumber
  P(\Sigma_a,t+\Delta t) =&& 
  \left[
   \sum_{b\neq a} P(\Sigma_b,t)w(\Sigma_a|\Sigma_b) + P(\Sigma_a,t)w(\Sigma_a|\Sigma_a) \right.\\
  &&\left.
  - \sum_{b\neq a} P(\Sigma_a,t)w(\Sigma_b|\Sigma_a)
  \right]\Delta t,
\end{eqnarray}
where $w(\Sigma_a|\Sigma_b)$ represents the transition probability from the $b$-th state to the $a$-th state in unit time.
The transition probability $w(\Sigma_a|\Sigma_b)$ obeys 
\begin{eqnarray}
 \sum_{\Sigma_a} w(\Sigma_a|\Sigma_b) = 1
  \qquad
  (\forall\, \Sigma_b).
\end{eqnarray}
For convenience, let ${\bf P}(t)$ be a vector-representation of probability distribution $\{P(\Sigma_a,t)\}$.
Then the master equation can be represented by
\begin{eqnarray}
 {\bf P}(t+\Delta t) = {\cal L}{\bf P}(t),
\end{eqnarray}
where ${\cal L}$ is the transition matrix whose elements are defined as
\begin{eqnarray}
 &&{\cal L}_{ba} := w(\Sigma_b|\Sigma_a) \Delta t,\\
 &&{\cal L}_{aa} := 1 - \sum_{b \neq a} {\cal L}_{ba} = 1 - \sum_{b \neq a} w(\Sigma_b|\Sigma_a)\Delta t.
\end{eqnarray}
Here the matrix ${\cal L}$ is a non-negative matrix and does not depend on time.
Then this time-evolution is the Markovian.

If the transition matrix ${\cal L}$ is prepared appropriately, which satisfies the detailed balance condition and the ergordicity, we can obtain the equilibrium probability distribution in the limit of infinite Monte Carlo step regardless of choice of the initial state because of the Perron-Frobenius theorem.

We can perform the Monte Carlo method easily as following process.
\begin{description}
 \item[Step 1] We prepare a initial state arbitrary.
 \item[Step 2] We choose a spin randomly.
 \item[Step 3] We calculate the molecular field at the chosen site in Step 2.
 The molecular field at the chosen site $i$ is defined as
 \begin{eqnarray}
  \label{STeq:molecularfield}
  h_i^{\rm (eff)} := \sum_j{}^\prime J_{ij} \sigma_j^z + h_i,
 \end{eqnarray}
 where the summation takes over the nearest neighbor sites of the $i$-th site.
 \item[Step 4] We flip the chosen spin in Step 2 according to a probability defined by some way.
 \item[Step 5] We continue from Step 2 to Step 4 until physical quantities such as magnetization converge.
\end{description}
In this Monte Carlo method, we only update the chosen single spin, and thus we refer to this method as single-spin-flip method.
There is an ambiguity how to define $w(\Sigma_a|\Sigma_b)$ in Step 4.
Here we explain two famous choices of $w(\Sigma_a|\Sigma_b)$ as follows.
Transition probability in the heat-bath method is given by
\begin{eqnarray}
 w_{\rm HB}(\sigma_i^z \to -\sigma_i^z) = 
  \frac{{\rm e}^{-\beta h_i^{\rm (eff)} \sigma_i^z}}{2\cosh (\beta h_i^{\rm (eff)})}.
\end{eqnarray}
Transition probability in the Metropolis method is given by
\begin{eqnarray}
 w_{\rm MP}(\sigma_i^z \to -\sigma_i^z) = 
  \begin{cases}
   1 \qquad & (h_i^{\rm (eff)} \sigma_i^z< 0)\\
   {\rm e}^{-2\beta h_i^{\rm (eff)}\sigma_i^z} \qquad & (h_i^{\rm (eff)} \sigma_i^z \ge 0)
  \end{cases}.
\end{eqnarray}
Since both two transition probabilities satisfy the detailed balance condition, the equilibrium state can be obtained definitely in the limit of infinite Monte Carlo step\footnote{Recently, the algorithm which does not use the detailed balance condition was proposed\cite{STSuwa-2010,STSuwa-2011}. It should be noted that the detailed balance condition is just a necessary condition. This novel algorithm is efficient for general spin systems.}.
It is important to select how to choice the transition probability since it is known that a couple of methods can sample states in an efficient fashion\cite{STSwendsen-1987,STWolff-1989,STHukushima-1996,STHarada-2004,STNakamura-2008,STMorita-2009,STSuwa-2010,STSuwa-2011}.

So far we considered the Monte Carlo method for systems where there is no off-diagonal matrix element.
To perform the Monte Carlo method, in a precise mathematical sense, we only have to know how to choice the basis or appropriate transformation so as to diagonalize the given Hamiltonian.
However, it is difficult to obtain equilibrium values of physical quantities of quantum systems, since we have to calculate the exponential of the given Hamiltonian ${\rm e}^{-\beta \hat{\cal H}}$ in general.
If we know all eigenvalues and the corresponding eigenvectors of the given Hamiltonian, we can easily calculate ${\rm e}^{-\beta \hat{\cal H}}$ by the unitary transformation which diagonalizes the Hamiltonian $\hat{\cal H}$.
In contrast, if we do not know all eigenvalues and eigenvectors, we have to calculate any power of the Hamiltonian $\hat{\cal H}^m$ since the matrix exponential is given by
\begin{eqnarray}
 \label{STeq:eqvalue_exponential}
 {\rm e}^{\hat{A}} = \sum_{m=0}^\infty \frac{1}{m!} \hat{A}^m.
\end{eqnarray}
It is difficult to calculate the matrix exponential in general.
Then we have to consider the following procedure in order to use the framework of the Monte Carlo method for quantum systems.

In many cases, the Hamiltonian of quantum systems can be represented as
\begin{eqnarray}
 \hat{{\cal H}} = \hat{{\cal H}}_{\rm c} + \hat{{\cal H}}_{\rm q}.
\end{eqnarray}
Hereafter we refer to $\hat{{\cal H}}_{\rm c}$ and $\hat{{\cal H}}_{\rm q}$ as classical Hamiltonian and quantum Hamiltonian, respectively.
The classical Hamiltonian $\hat{{\cal H}}_{\rm c}$ is a diagonal matrix.
Here we assume that $\hat{{\cal H}}_{\rm q}$ can be easily diagonalized\footnote{
This fact does not seem to be general.
However we can prepare the matrices which can be easily diagonalized by the decomposition as $\hat{{\cal H}}_{\rm q} = \sum_{\ell} \hat{{\cal H}}_{\rm q}^{(\ell)}$ in many cases.
}.
This is a key of the quantum Monte Carlo method as will be shown later.
Since $\hat{{\cal H}}_{\rm c}$ and $\hat{{\cal H}}_{\rm q}$ cannot commute in general: $[\hat{{\cal H}}_{\rm c},\hat{{\cal H}}_{\rm q}]\neq 0$, then ${\rm e}^{-\beta {\hat {\cal H}}} \neq {\rm e}^{-\beta\hat{{\cal H}}_{\rm c}}{\rm e}^{-\beta\hat{{\cal H}}_{\rm q}}$.
We decompose the matrix exponential by introducing large integer $m$,
\begin{eqnarray}
 \nonumber 
  \exp\left( {-\frac{\beta}{m}\hat{{\cal H}}}\right) 
  &&= \exp\left[ -\frac{\beta}{m} (\hat{{\cal H}}_{\rm c} + \hat{{\cal H}}_{\rm q})\right]\\
 \label{STeq:Trotter}
 &&=\exp\left( -\frac{\beta}{m} \hat{{\cal H}}_{\rm c} \right)
  \exp\left( -\frac{\beta}{m} \hat{{\cal H}}_{\rm q} \right)
  + {\cal O}\left( \left( \frac{\beta}{m}\right)^2\right).
\end{eqnarray}
This is a concrete representation of the Trotter formula\cite{STTrotter-1959}.
From now on, we refer to $m$ as Trotter number.
By using this relation, we can perform the Monte Carlo method for quantum systems.
To illustrate it, we consider the Ising model with longitudinal and transverse magnetic fields.
The considered Hamiltonian is given as
\begin{eqnarray}
 &&\hat{{\cal H}} = 
  - \sum_{\langle i,j \rangle} J_{ij} \hat{\sigma}_i^z \hat{\sigma}_j^z
  - \sum_{i=1}^N h_i^z \hat{\sigma}_i^z
  - \Gamma \sum_{i=1}^N \hat{\sigma}_i^x
  = \hat{{\cal H}}_{\rm c} + \hat{{\cal H}}_{\rm q},\\
 &&\hat{{\cal H}}_{\rm c} := 
  - \sum_{\langle i,j \rangle} J_{ij} \hat{\sigma}_i^z \hat{\sigma}_j^z
  - \sum_{i=1}^N h_i^z \hat{\sigma}_i^z,
  \qquad
  \hat{{\cal H}}_{\rm q} := 
  - \Gamma \sum_{i=1}^N \hat{\sigma}_i^x,
\end{eqnarray}
where optimization problems often can be expressed by this classical Hamiltonian $\hat{{\cal H}}_{\rm c}$.
The partition function of the Hamiltonian at temperature $T(=\beta^{-1})$ is given by
\begin{eqnarray}
   \label{STeq:bef_TSdecompose_partitionfunction}
 Z = {\rm Tr}\, {\rm e}^{-\beta \hat{{\cal H}}} 
  = \sum_\Sigma \Braket{\Sigma | {\rm e}^{-\beta(\hat{{\cal H}}_{\rm c}+\hat{{\cal H}}_{\rm q})}|\Sigma}.
\end{eqnarray}
Using Eq.~(\ref{STeq:Trotter}) we obtain
\begin{eqnarray}
\nonumber 
Z &&= \lim_{m\to \infty}\sum_{\{\Sigma_k\}, \{\Sigma'_k\}}
  \Braket{\Sigma_1 | {\rm e}^{-\beta\hat{{\cal H}}_{\rm c}/m} | \Sigma_1'}
  \Braket{\Sigma_1' | {\rm e}^{-\beta\hat{{\cal H}}_{\rm q}/m} | \Sigma_2}\\
 \nonumber 
  &&\times
  \Braket{\Sigma_2 | {\rm e}^{-\beta\hat{{\cal H}}_{\rm c}/m} | \Sigma_2'}
  \Braket{\Sigma_2' | {\rm e}^{-\beta\hat{{\cal H}}_{\rm q}/m} | \Sigma_3}\\
 \nonumber 
  &&
  \times \cdots \\ 
  &&
   \label{STeq:TSdecompose_partitionfunction}
  \times
  \Braket{\Sigma_m | {\rm e}^{-\beta\hat{{\cal H}}_{\rm c}/m} | \Sigma_m'}
  \Braket{\Sigma_m' | {\rm e}^{-\beta\hat{{\cal H}}_{\rm q}/m} | \Sigma_1},
\end{eqnarray}
where $\ket{\Sigma_k}$ represents the direct-product space of $N$ spins:
\begin{eqnarray}
 \ket{\Sigma_k} := \ket{\sigma_{1,k}^z} \otimes \ket{\sigma_{2,k}^z} \otimes \cdots \ket{\sigma_{N,k}^z},
\end{eqnarray}
where the first and the second subscripts of $\ket{\sigma_{i,k}^z}$ indicate coordinates of the real space and the Trotter axis, respectively.
Here $\ket{\sigma_{i,k}^z}=\ket{\uparrow}$ or $\ket{\downarrow}$.
Equation~(\ref{STeq:bef_TSdecompose_partitionfunction}) consists of two elements $\braket{\Sigma_k | {\rm e}^{-\beta\hat{{\cal H}}_{\rm c}/m} | \Sigma_k'}$ and $\braket{\Sigma_k' | {\rm e}^{-\beta\hat{{\cal H}}_{\rm q}/m} | \Sigma_{k+1}}$.
Since the classical Hamiltonian $\hat{{\cal H}}_{\rm c}$ is a diagonal matrix, the former is easily calculated:
\begin{eqnarray}
 \nonumber
 &&\Braket{\Sigma_k | {\rm e}^{-\beta\hat{{\cal H}}_{\rm c}/m} | \Sigma_k'}\\
 && = \exp
  \left[
   \frac{\beta}{m} 
   \left(
    \sum_{\langle i,j \rangle} J_{ij} \sigma_{i,k}^z \sigma_{j,k}^z + \sum_{i=1}^N h_i \sigma_{i,k}^z
   \right)
   \right]
  \prod_{i=1}^N \delta(\sigma_{i,k}^z,\sigma_{i,k}'^z),
\end{eqnarray}
where $\sigma_{i,k}^z = \pm1$.
On the other hand, the latter $\Braket{\Sigma_k' | {\rm e}^{-\beta\hat{{\cal H}}_{\rm q}/m} | \Sigma_{k+1}}$ is calculated as
\begin{eqnarray}
 \nonumber
 &&\Braket{\Sigma_k' | {\rm e}^{-\beta\hat{{\cal H}}_{\rm q}/m} | \Sigma_{k+1}}\\
 && = 
  \left[
   \frac{1}{2} \sinh \left( \frac{2\beta\Gamma}{m}\right)
  \right]^{N/2}
  \exp
  \left[
   \frac{1}{2} \ln \coth
   \left(
    \frac{\beta\Gamma}{m}\sum_{i=1}^N
    \sigma_{i,k}'^z \sigma_{i,k+1}^z
   \right)
  \right].
\end{eqnarray}
Then the partition function given by Eq.~(\ref{STeq:TSdecompose_partitionfunction}) can be represented as
\begin{eqnarray}
 \nonumber 
  Z =
  \lim_{m\to \infty} A \sum_{\{\sigma_{i,k}^z = \pm 1\}}
  \exp
  &&\left\{
     \sum_{k=1}^m
     \left[
     \sum_{\langle i,j \rangle}
     \left(
      \frac{\beta J_{ij}}{m} \sigma_{i,k}^z \sigma_{j,k}^z
     \right)
     + \sum_{i=1}^N \frac{\beta h_i}{m} \sigma_{i,k}^z
    \right.\right.\\
 &&\left.\left.
   + \sum_{i=1}^N \frac{1}{2} \ln \coth 
   \left(
    \frac{\beta\Gamma}{m}
    \right)
    \sigma_{i,k}^z \sigma_{i,k+1}^z
   \right]
   \right\},
\end{eqnarray}
where $A$ is just a parameter which does not affect physical quantities.
It should be noted that the partition function of the $d$-dimensional Ising model with transverse field $\hat{\cal H}$ is equivalent to that of the $(d+1)$-dimensional Ising model {\it without} transverse field ${\cal H}_{\rm eff}$ which is given by
\begin{align}
{\cal H}_{\rm eff} = 
    & -\sum_{\langle i,j \rangle} \sum_{k=1}^m
      \frac{J_{ij}}{m} \sigma_{i,k}^z \sigma_{j,k}^z
           - \sum_{i=1}^N \sum_{k=1}^m \frac{h_i}{m} \sigma_{i,k}^z \notag \\
    &- \frac{1}{\beta} \sum_{i=1}^N \sum_{k=1}^m \frac{1}{2} \ln \coth 
    \left(
    \frac{\beta\Gamma}{m}
    \right)
    \sigma_{i,k}^z \sigma_{i,k+1}^z.
\end{align}
The coefficient of the third term of RHS is always negative,
and thus the interaction along the Trotter axis is always ferromagnetic.
This ferromagnetic interaction becomes strong as the value of $\Gamma$ decreases.
This is called the Suzuki-Trotter decomposition\cite{STTrotter-1959,STSuzuki-1976}.

So far we explained the Monte Carlo method as a tool for obtaining the equilibrium state.
However we can also use this method to investigate stochastic dynamics of strongly correlated systems, since the Monte Carlo method is originally based on the master equation.
In terms of optimization problem, our purpose is to obtain the ground state of the given Hamiltonian.
Then we decrease transverse field gradually and obtain a solution.
There are many Monte Carlo studies in which the quantum annealing succeeds to obtain a better solution than that by the simulated annealing\cite{STKadowaki-1998b,STKurihara-2009,STTanaka-2011a,STKadowaki-1998a,STSantoro-2002,STDas-2005,STDas-2008}.


\subsection{Deterministic Method Based on Mean-Field Approximation}
\label{STsubsec:MF}

In the previous section, we considered the Monte Carlo method in which time-evolution is treated as stochastic dynamics.
In this section, on the other hand, we explain a deterministic method based on mean-field approximation according to Refs.~[\refcite{STTanaka-2000,STTanaka-2002}].
Before we consider the quantum annealing based on the mean-field approximation, we treat the Ising model with random interactions and site-dependent longitudinal fields given by
\begin{eqnarray}
 {\cal H}_{\rm Ising} 
  = - \sum_{\langle i,j \rangle} J_{ij} \sigma_i^z \sigma_j^z
  - \sum_{i=1}^N h_i \sigma_i^z.
\end{eqnarray}
When the transverse field is absent, the molecular field of the $i$-th spin is given by Eq.~(\ref{STeq:molecularfield}).
Then an equation which determines expectation value of the $i$-th spin at temperature $T(=\beta^{-1})$ is given by
\begin{eqnarray}
 m_i^z = 
  \frac{
  {\rm e}^{\beta h_i^{\rm (eff)}} - {\rm e}^{-\beta h_i^{\rm (eff)}}
  }{
  {\rm e}^{\beta h_i^{\rm (eff)}} + {\rm e}^{-\beta h_i^{\rm (eff)}}  
  }
  = \tanh (\beta h_i^{\rm (eff)}).
\end{eqnarray}
In the mean-field level, we approximate that the state $\sigma_j^z$ is equal to the expectation value $m_j^z$ in Eq.~(\ref{STeq:molecularfield}), and we obtain
\begin{eqnarray}
 m_i^z = \tanh 
  \left[
   \beta \left( \sum_j{}^\prime J_{ij} m_j^z + h_i \right)
  \right],
\end{eqnarray}
which is often called self-consistent equation.

We can obtain equilibrium value in the mean-field level by iterating the following equation until convergence:
\begin{eqnarray}
 m_i^z(t+1) = \tanh (\beta h_i^{\rm (eff)}(t)),
  \qquad
  h_i^{\rm (eff)}(t) = \sum_{j}{}^\prime J_{ij} m_j^z(t) + h_i.
\end{eqnarray}
In order to judge the convergence, we introduce a distance which represents difference between the state at $t$-th step and that at $(t+1)$-th step as follows:
\begin{eqnarray}
 d(t) := \frac{1}{N} \sum_{i=1}^N
  \left|
   m_i^z(t+1) - m_i^z(t)
  \right|.
\end{eqnarray}
When the quantity $d(t)$ is less than a given small value (typically $\sim 10^{-8}$ or more smaller value), we judge that the calculation is converged.
We summarize this method:
\begin{description}
 \item[Step 1] We prepare a initial state arbitrary.
 \item[Step 2] We choose a spin randomly.
 \item[Step 3] We calculate the molecular field given by Eq.~(\ref{STeq:molecularfield}) at the chosen site in Step 2.
 \item[Step 4] We change the value of the chosen spin in Step 2 according to the obtained molecular field in Step 3.
 \item[Step 5] We continue from Step 2 to Step 4 until the distance $d(t)$ converges to small value.
\end{description}
The differences between the Monte Carlo method and this method are Step 4 and Step 5.
We can perform the simulated annealing by decreasing temperature and using the state obtained in Step 5 as the initial state in Step 1 at the time changing temperature\footnote{If we want to decrease temperature rapidly, we choose not so small value for judgement of convergence.}.

Next we explain a quantum version of this method.
Here we apply transverse field as a quantum field.
We consider the Hamiltonian given by
\begin{eqnarray}
 \hat{\cal H} = 
  - \sum_{\langle i,j \rangle} J_{ij} \hat{\sigma}_i^z \hat{\sigma}_j^z
  - \sum_{i=1}^N h_i \hat{\sigma}_i^z
  - \Gamma \sum_{i=1}^N \hat{\sigma}_i^x.
\end{eqnarray}
The density matrix of the equilibrium state is 
\begin{eqnarray}
 \hat{\rho} = 
  \frac{
  \exp (-\beta \hat{{\cal H}})
  }{
  {\rm Tr}\, \exp(-\beta \hat{{\cal H}})
  } = 
  \frac{
  \sum_{n=1}^{2^N} \exp (-\beta \epsilon_n) \ket{\lambda_n}\bra{\lambda_n}
  }{
  \sum_{n=1}^{2^N} \exp (-\beta \epsilon_n)
  },
\end{eqnarray}
where $\epsilon_n$ and $\ket{\lambda_n}$ denote the $n$-th eigenenergy and the corresponding eigenvector.
The density matrix satisfies the variational principle that minimizes free energy:
\begin{eqnarray}
 F = \min_{\hat{\rho}} 
  \left[
   {\rm Tr}\, (\hat{{\cal H}} + \beta^{-1}\ln \hat{\rho})\hat{\rho}
  \right],
\end{eqnarray}
where the logarithm of the matrix is defined by the series expansion as well as the definition of the matrix exponential (see Eq.~(\ref{STeq:eqvalue_exponential})).
Since it is difficult to obtain the density matrix, we have to consider alternative strategy as follows.

A reduced density matrix is defined as
\begin{eqnarray}
 \hat{\rho}_i := {\rm Tr}'\, \hat{\rho}
  = \frac{1}{2}
  \left(
   \hat{I} + m_i^z \hat{\sigma}^z + m_i^x \hat{\sigma}^x
  \right),
\end{eqnarray}
where ${\rm Tr}'$ indicates trace over spin states except the $i$-th spin.
The values $m_i^z$ and $m_i^x$ are calculated by
\begin{eqnarray}
 m_i^z = {\rm Tr}\, (\hat{\sigma}_i^z \hat{\rho}),
  \qquad
  m_i^x = {\rm Tr}\, (\hat{\sigma}_i^x \hat{\rho}).
\end{eqnarray}
The reduced density matrix satisfies the following relations:
\begin{eqnarray}
 {\rm Tr}\, (\hat{\rho}_i) = 1,
  \qquad
  {\rm Tr}\, (\hat{\sigma}_i^z \hat{\rho}_i) = m_i^z,
  \qquad
  {\rm Tr}\, (\hat{\sigma}_i^x \hat{\rho}_i) = m_i^x.
\end{eqnarray}
Here we assume that the density matrix can be represented by direct products of the reduced density matrices:
\begin{eqnarray}
 \hat{\rho} \simeq \prod_{i=1}^N \hat{\rho}_i,
\end{eqnarray}
which is mean-field approximation (in other words, decoupling approximation).
Then, the free energy is expressed as
\begin{eqnarray}
 &&F \simeq \min_{\{\hat{\rho}_i\}} {\cal F}(\{\hat{\rho}_i\}),\\
\nonumber 
 &&{\cal F}(\{\hat{\rho}_i\})
 = -\sum_{\langle i,j \rangle} J_{ij} m_i^z m_j^z - \sum_{i=1}^N h_i m_i^z - \Gamma \sum_{i=1}^N m_i^x\\
 &&\ \ \ \ \ \ \ \ \ \ \ \ \ + \beta^{-1} \sum_{i=1}^N {\rm Tr}\, ( \hat{\rho}_i \ln \hat{\rho}_i).
\end{eqnarray}
From the variation of ${\cal F}(\{\hat{\rho}_i\})$ under the normalization condition, we obtain the following relations:
\begin{eqnarray}
 &&\hat{\rho}_i = 
  \frac{
  \exp(-\beta \hat{\cal H}_i)
  }{
  {\rm Tr}\, [ \exp (-\beta \hat{\cal H}_i)]
  },\\
 &&
  \hat{\cal H}_i = 
  \left(
   \begin{array}{cc}
   -h_i - \sum_{j}' J_{ij} m_j^z &-\Gamma \\
   -\Gamma & +h_i + \sum_{j}' J_{ij} m_j^z
   \end{array}
  \right).
\end{eqnarray}
Then the reduced density matrix is represented by using the $n$-th ($n=1,2$) eigenvalues $\epsilon_n^{(i)}$ and the corresponding eigenvectors $\ket{\lambda_n^{(i)}}$ of $\hat{\cal H}_i$:
\begin{eqnarray}
 \hat{\rho}_i 
  = \frac{
  \exp(-\beta\epsilon_1^{(i)}) \ket{\lambda_1^{(i)}}\bra{\lambda_1^{(i)}} + 
  \exp(-\beta\epsilon_2^{(i)}) \ket{\lambda_2^{(i)}}\bra{\lambda_2^{(i)}} 
  }{
  \exp(-\beta\epsilon_1^{(i)}) + \exp(-\beta\epsilon_2^{(i)})
  }.
\end{eqnarray}
We can also obtain the equilibrium values of physical quantities as well as the case for $\Gamma=0$:
\begin{eqnarray}
 &&m_i^z(t+1) = {\rm Tr} (\hat{\sigma}_i^z \hat{\rho}_i (t)),
  \qquad
 m_i^x(t+1) = {\rm Tr} (\hat{\sigma}_i^x \hat{\rho}_i (t)),\\
 && \hat{\rho}_i (t) =  
  \frac{
  \exp(-\beta \hat{\cal H}_i(t))
  }{
  {\rm Tr}\, \exp (-\beta \hat{\cal H}_i(t))
  },\\
  &&\hat{\cal H}_i(t) = 
  \left(
   \begin{array}{cc}
    -h_i - \sum_{j}' J_{ij} m_j^z(t) &-\Gamma \\
   -\Gamma & +h_i + \sum_{j}' J_{ij} m_j^z(t)
   \end{array}
  \right).
\end{eqnarray}
We continue the above self-consistent equation until the following distance converges:
\begin{eqnarray}
 d(t) := \frac{1}{2N} \sum_{i=1}^N
  \left(  
   \left| m_i^z(t+1) - m_i^z(t) \right| +
   \left| m_i^x(t+1) - m_i^x(t) \right|
  \right).
\end{eqnarray}

If the temperature is zero, the reduced density matrix should be
\begin{eqnarray}
 \hat{\rho}_i = \ket{\lambda_1^{(i)}}\bra{\lambda_1^{(i)}},
\end{eqnarray}
where we consider the case for $\epsilon_1^{(i)} < \epsilon_2^{(i)}$.
Note that if and only if $-h_i - \sum_{j}' J_{ij} m_j^z =\Gamma = 0$, $\epsilon_1^{(i)} = \epsilon_2^{(i)}$ is satisfied.
Then if we perform the quantum annealing at $T=0$, we have to know only the ground state of the local Hamiltonian $\hat{\cal H}_i$.
The procedure is the same as the case for finite temperature.
By using the method, we can obtain a better solution than that obtained by the simulated annealing for some optimization problems.
Recently, other type of implementation method based on mean-field approximation was proposed\cite{STSato-2009}.
The method is a quantum version of the variational Bayes inference\cite{STAttias-1999}.
We can also obtain a better solution than the conventional variational Bayes inference.


\subsection{Real-Time Dynamics}
\label{STsubsec:RT}

In Sec.~\ref{STsubsec:MC} and Sec.~\ref{STsubsec:MF}, we considered artificial time-development rules such as the Markov chain Monte Carlo method and mean-field dynamics.
In this section, we explain real-time dynamics which is expressed by the time-dependent Schr\"odinger equation:
\begin{eqnarray}
 i \frac{\partial}{\partial t}\ket{\psi(t)} = \hat{\cal H}(t) \ket{\psi(t)},
\end{eqnarray}
where $\hat{{\cal H}}(t)$ and $\ket{\psi(t)}$ denote the time-dependent Hamiltonian and the wave function at time $t$, respectively.
The solution of this equation is given by
\begin{eqnarray}
 \label{STeq:Schrodingerequation}
 \ket{\psi(t)} = \exp 
  \left[
   -i \int_0^t \hat{\cal H}(t'){\rm d}t'
  \right]
  \ket{\psi(t=0)}.
\end{eqnarray}
If we use the time-dependent Hamiltonian including time-dependent quantum field, we can perform the quantum annealing by decreasing the quantum field gradually.
To obtain the solution, it is necessary to decide the initial state for Eq.~(\ref{STeq:Schrodingerequation}).
Since our purpose is to obtain the ground state of the given Hamiltonian which represents the optimization problem, we have no way to know the preferable initial state that leads to the ground state definitely in the adiabatic limit.
However, in general, we often use a ``trivial state'' as the initial state.
Actually, it goes well in many cases.
For instance, when we consider the Ising model with time-dependent transverse field which is given by
\begin{eqnarray}
 \hat{\cal H}(t) = - \sum_{i,j} J_{ij} \hat{\sigma}_i^z \hat{\sigma}_j^z 
  - \Gamma(t) \sum_{i=1}^N \hat{\sigma}_i^x,
\end{eqnarray}
we set the ground state for large $\Gamma$ as the initial state, hence the initial state is set as 
\begin{eqnarray}
 \ket{\psi(t=0)} = \ket{\rightarrow \rightarrow \cdots \rightarrow},
\end{eqnarray}
where $\ket{\rightarrow}$ denotes the eigenstate of $\hat{\sigma}^x$:
\begin{eqnarray}
 \ket{\rightarrow} := \frac{1}{\sqrt{2}} ( \ket{\uparrow} + \ket{\downarrow}).
\end{eqnarray}

\begin{figure}[t]
 \begin{center}
 \psfig{file=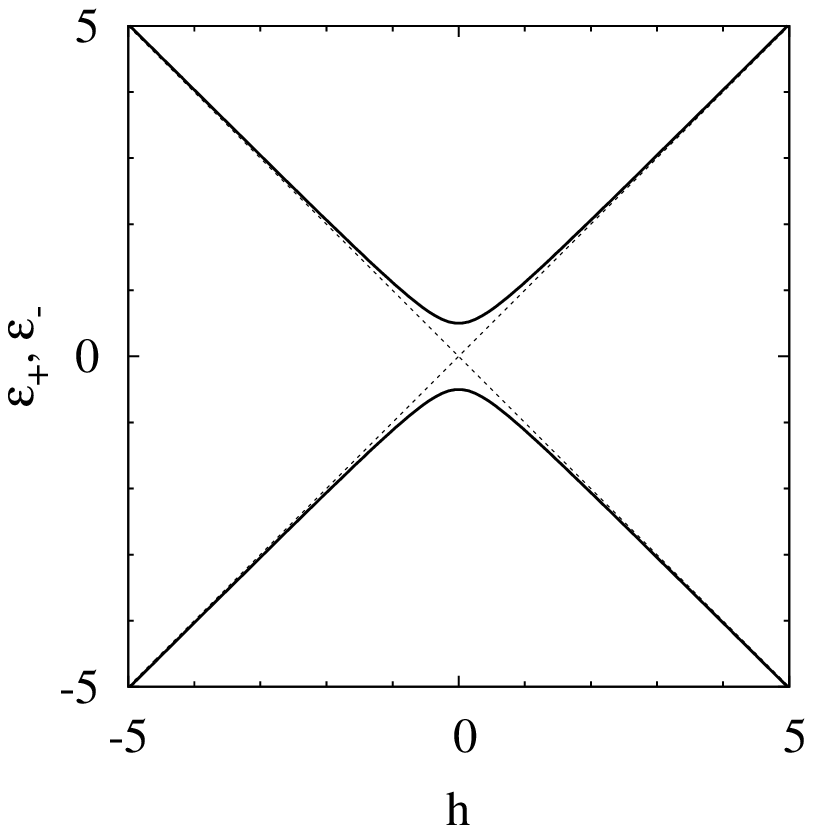,scale=0.6}
 \psfig{file=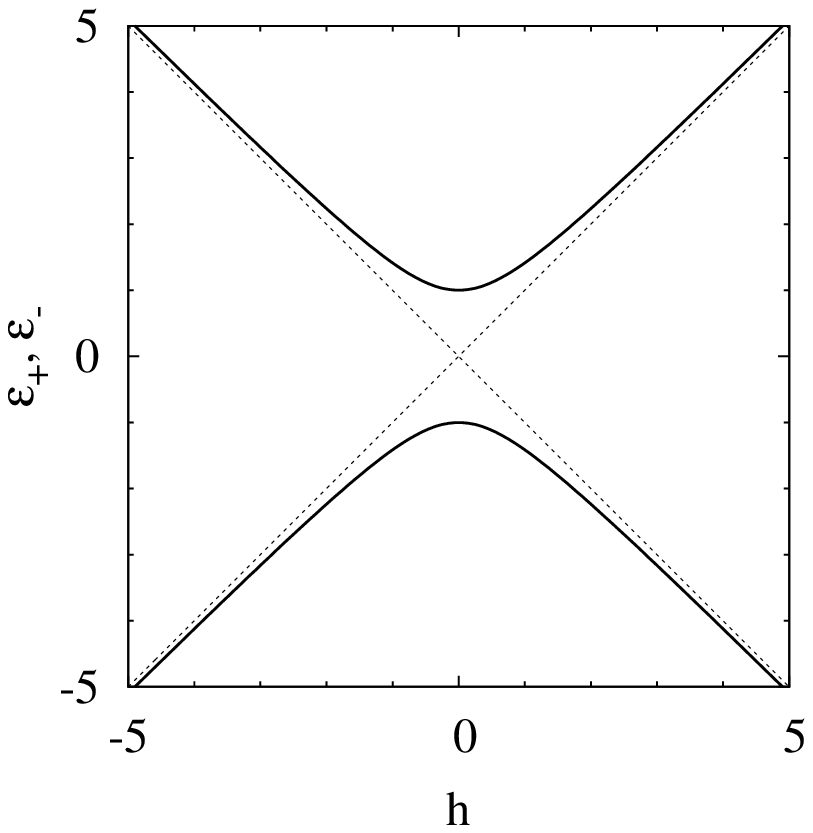,scale=0.6}
 \end{center}
 \caption{
 Eigenenergies of the single spin system under longitudinal and transverse magnetic fields for $\Gamma=0.5$ (left panel) and $\Gamma=1$ (right panel).
 The dotted lines represent eigenenergies for $\Gamma=0$.
 }
 \label{STfig:singlespinhg}
\end{figure}

In real-time dynamics, in order to obtain the ground state by using given initial condition, it is important whether there is level crossing.
If there is no level crossing, the system can necessarily reach the ground state by the quantum annealing in the adiabatic limit.
To show this fact, we first consider a single spin system under time-dependent longitudinal magnetic field.
The Hamiltonian is given by
\begin{eqnarray}
 \hat{\cal H}_{\rm single}(t) = -h(t) \hat{\sigma}^z = 
  \left(
   \begin{array}{cc}
    -h(t) & 0\\
    0  & h(t)\\
   \end{array}
  \right).
\end{eqnarray}
Suppose we set $\ket{\psi(0)} = \ket{\downarrow}$ as the initial state.
For arbitrary sweeping schedules, the state at arbitrary positive $t$ is obtained by
\begin{eqnarray}
 \ket{\psi(t)} = \exp 
  \left[
   -i \int_0^t \hat{\cal H}_{\rm single}(t'){\rm d}t'
  \right]
  \ket{\psi(0)}
  = \ket{\downarrow}.
\end{eqnarray}
This is because the state $\ket{\downarrow}$ is the eigenstate of the instantaneous Hamiltonian for arbitrary time $t$.
In general, when there is a good quantum number and the initial state is set to be the corresponding eigenstate, the good quantum number is conserved.
Then when we perform the quantum annealing method based on the real-time dynamics, we should take care of the symmetries of the considered Hamiltonian.
From this, we can obtain the ground state of the considered system in the adiabatic limit if there is no level crossing.
In practice, however, since we change magnetic field with finite speed, a nonadiabatic transition is inevitable.
To show this fact, we consider a single spin system under longitudinal and transverse magnetic fields.
The Hamiltonian of this system is given by
\begin{eqnarray}
 \hat{\cal H}_{\rm single} = - h \hat{\sigma}^z - \Gamma \hat{\sigma}^x = 
  \left(
   \begin{array}{cc}
    -h & -\Gamma \\
    -\Gamma & h
   \end{array}
  \right).
\end{eqnarray}
Since the eigenenergies are $\epsilon_{\pm} = \pm \sqrt{h^2 + \Gamma^2}$, the smallest value of the energy difference between the ground state and the excited state is $2\Gamma$ at $h=0$ as shown in Fig.~\ref{STfig:singlespinhg}.

Suppose we consider the single spin system under time-dependent longitudinal magnetic field and fixed transverse magnetic field.
The Hamiltonian is given by
\begin{eqnarray}
 \hat{\cal H}_{\rm single} (t) = - h(t) \hat{\sigma}^z - \Gamma \hat{\sigma}^x = 
  \left(
   \begin{array}{cc}
    -vt & -\Gamma \\
    -\Gamma & vt
   \end{array}
  \right),
\end{eqnarray}
where we adopt $h(t) = vt$ as time-dependent longitudinal field.
Here we set $t=-\infty$ as the initial time.
The initial state is set to be the ground state of the Hamiltonian at the initial time $\ket{\psi(t=-\infty)}=\ket{\downarrow}$.
The ground state at $t=+\infty$ in the adiabatic limit is $\ket{\psi^{\rm (ad)}(t=+\infty)}=\ket{\uparrow}$.
Then a characteristic value which represents the nature of this dynamics is a probability of staying in the ground state at $t=+\infty$ which is defined by
\begin{eqnarray}
\nonumber 
 P_{\rm stay} &&= \Braket{\psi^{\rm (ad)}(t=+\infty) | \exp 
 \left[
  -i \int_{-\infty}^{+\infty} \hat{\cal H}_{\rm single}(t'){\rm d}t'
 \right] | \psi(t=-\infty)}\\
 &&= \Braket{\uparrow | \exp 
 \left[
  -i \int_{-\infty}^{+\infty} \hat{\cal H}_{\rm single}(t'){\rm d}t'
 \right] | \downarrow}.
\end{eqnarray}
The probability of staying in the ground state should depend on the sweeping speed $v$ and the characteristic energy gap and can be obtained by the Landau-Zener-St\"uckelberg formula\cite{STLandau-1932,STZener-1932,STStuckelberg-1932}:
\begin{eqnarray}
 P_{\rm stay} = 1 - \exp
  \left[
   -\frac{
   \pi(\Delta E)^2
   }{
   4v\Delta m
   }
  \right],
\end{eqnarray}
where $\Delta E$ and $\Delta m$ represent the energy gap at the avoided level-crossing point and the difference of the magnetizations in the adiabatic limit, respectively.
In this case $\Delta E = 2\Gamma$ and $\Delta m = 2$.

In many cases, typical shape of energy structure can be approximated by simple systems such as the single spin system.
Then the knowledge of the simple transitions such as the Landau-Zener-St\"ukelberg transition and the Rosen-Zener transition\cite{STRosen-1932b} is useful to analyze the efficiency of the quantum annealing based on the real-time dynamics.

\subsection{Experiments}

Transverse field response of the Ising model has been also established in experimentally\cite{STChakrabarti-1996,STTrammel-1960,STCooke-1962,STStout-1962,STMoruzzi-1963,STNarath-1966,STWielinga-1969,STWolf-1971,STWu-1991,STWu-1993}.
A dipolar-coupled disordered magnet LiHo$_x$Y$_{1-x}$F$_4$ has easy-axis anisotropy and can be represented by the Ising model\cite{STReich-1990,STRosenbaum-1996}.
If we apply the longitudinal magnetic field (in other words, the magnetic field is parallel to the easy-axis), phase transition does not take place\cite{STReich-1986,STMydosh-1993}.
However, when we apply the transverse magnetic field (in other words, the magnetic field is perpendicular to the easy-axis), phase transitions occur and interesting dynamical properties shown in Ref.[~\refcite{STBrooke-1999}] were observed.
In the phase diagram of this material, there are three phases.
The ferromagnetic phase appears at intermediate temperature and low transverse magnetic field, whereas at low temperature and low transverse magnetic field, the glassy critical phase\cite{STBak-1987} appears.
The paramagnetic phase exists at the other region.
The glassy critical phase exhibits slow relaxation in general.
It found that the characteristic relaxation time obtained by ac field susceptibility for quantum cooling in which we decrease transverse field after temperature is decreased is lower than that for temperature cooling case\cite{STBrooke-1999}.
From this result, it has been expected that the effect of the quantum fluctuation helps us to obtain the best solution of the optimization problem.
\section{Optimization Problems}
\label{STsec:optimization_problems}

Optimization problems are defined by composition elements of the considered problem and real-valued cost/gain function.
They are problems to obtain the best solution such that the cost/gain function takes the minimum/maximum value.
In general, the number of candidate solutions increases exponentially with the number of composition elements in optimization problems.
Although we can obtain the best solution by a brute force in principle, it is difficult to obtain the best solution by such a naive method in practice.
Then we have to invent an innovative method for obtaining the best solution in a practical time and limited computational resource.
Optimization problems can be expressed by the Ising model in many cases.
Once optimization problems are mapped onto the Ising model, we can use methods that have been considered in statistical physics and computational physics such as the quantum annealing. 

In the anterior half of this section, we explain the correspondence between the Ising model and the traveling salesman problem which is one of famous optimization problems.
We demonstrate the quantum annealing based on the quantum Monte Carlo simulation for this problem.
In the posterior half, we explain the clustering problem as the example expressed by the Potts model which is a straightforward extension of the Ising model.

\subsection{Traveling Salesman Problem}
\label{STsec:tsp}

In this section, we consider the traveling salesman problem which is one of famous optimization problems.
The setup of the traveling salesman problem is as follows:
\begin{itemize}
 \item There are $N$ cities.
 \item We move from the $i$-th city to the $j$-th city where the distance between them is $\ell_{i,j}$.
 \item We can pass through a city only once.
 \item We return the initial city after we pass through all the cities.
\end{itemize}
The traveling salesman problem is to find the minimum path under above conditions.
The length of a path is given by
\begin{eqnarray}
 L := \sum_{a=1}^N \ell_{c_a,c_{a+1}},
\end{eqnarray}
where $c_a$ denotes the city where we pass through at the $a$-th step.
In the traveling salesman problem, the length of a path is a cost function.
From the fourth condition, the following relation should be satisfied:
\begin{eqnarray}
 c_{N+1} = c_1.
\end{eqnarray}
In terms of mathematics, the traveling salesman problem is to find $\{c_a\}_{a=1}^N$ so as to minimize the path $L$ under the above four conditions.

\begin{figure}[b]
 \begin{center}
 \psfig{file=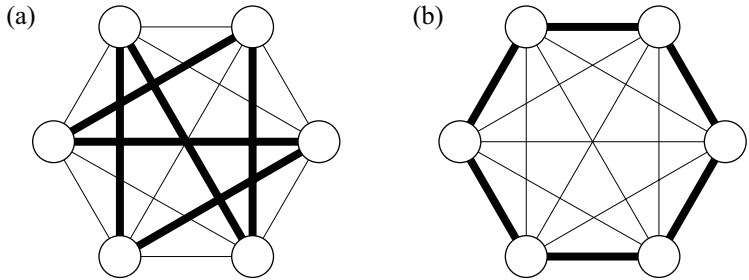,scale=1.0}
 \end{center}
 \caption{
 Traveling salesman problem for $N=6$.
 Thin lines and thick lines denote the permitted paths and selected paths, respectively.
 (a) Bad solution.
 (b) The best solution in which the length of the path is minimum.
 }
 \label{STfig:tspexample}
\end{figure}

If the number of cities $N$ is small, it is easy to obtain the shortest path by a brute force.
We can easily find the best solution of the traveling salesman problem for $N=6$ shown in Fig.~\ref{STfig:tspexample}.
Figure~\ref{STfig:tspexample} (a) and (b) represent a bad solution and the best solution where the length of the path $L$ is minimum, respectively.
As the number of cities increases, the traveling salesman problem becomes seriously difficult since the number of candidate solutions is $(N-1)!/2$.
Then if we want to deal with the traveling salesman problem with large $N$, we have to adopt smart and easy practical methods such as the simulated annealing instead of a brute force.
To use the simulated annealing, we map the traveling salesman problem onto the Ising model with a couple of constraints as follows.

We consider $N \times N$ two-dimensional lattice.
Let $n_{i,a}$ be the microscopic state which represents the state at the $i$-th city at the $a$-th step.
The value of $n_{i,a}$ can be taken either $0$ or $1$.
If we pass through the $i$-th city at the $a$-th step, $n_{i,a}$ is unity whereas $n_{i,a}=0$ if we do not pass through the $i$-th city at the $a$-th step.
The third condition can be represented by
\begin{eqnarray}
 \label{STeq:constraint_tsp_1}
 \sum_{a=1}^N n_{i,a} = 1 \qquad ({\rm for}\,\, \forall i).
\end{eqnarray}
Furthermore, since it is obvious that we can pass through only one city at the $a$-th step, this constraint is expressed by
\begin{eqnarray}
 \label{STeq:constraint_tsp_2}
 \sum_{i=1}^N n_{i,a} = 1 \qquad ({\rm for}\,\, \forall a).
\end{eqnarray}
Then the length of the path $L$ can be rewritten as
\begin{eqnarray}
 L = \sum_{a=1}^N \sum_{i,j} \ell_{i,j} n_{i,a} n_{j,a+1}
  = \frac{1}{4}
  \sum_{a=1}^N \sum_{i,j} \ell_{i,j} \sigma_{i,a}^z \sigma_{j,a+1}^z + {\rm const.},
\end{eqnarray}
where the Ising spin variable $\sigma_{i,a}^z = \pm 1$ is defined by 
\begin{eqnarray}
 \sigma_{i,a}^z := 2 n_{i,a} - 1.
\end{eqnarray}
Here we used the following relation derived by Eqs.~(\ref{STeq:constraint_tsp_1}) and (\ref{STeq:constraint_tsp_2}):
\begin{eqnarray}
 \sum_{a=1}^N \sum_{i,j} \ell_{i,j} \sigma_{i,a}^z = {\rm const.}
\end{eqnarray}
Then the length of the path can be represented by the Ising spin Hamiltonian on $N \times N$ two-dimensional lattice.
In general, it is difficult to obtain the stable state of the Ising model with some constraints regarded as some kind of frustration which will be shown in Sec.~\ref{STsec:Frustration}.

\subsubsection{Monte Carlo Method}

\begin{figure}[t]
 \begin{center} 
 \psfig{file=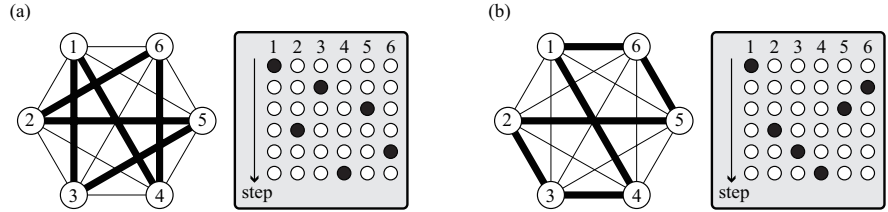,scale=0.9}
 \end{center}
 \caption{
 The simplest way of flipping method in traveling salesman problem.
 Transition between the state depicted in (a) and that depicted in (b) occurs.
 In this case, $i=3$, $j=6$, $a=2$, and $a'=5$.
 }
 \label{STfig:TSP_protocol}
\end{figure}

We explain how to implement the Monte Carlo method in the traveling salesman problem.
We cannot use the single-spin-flip method which was explained in Sec.~\ref{STsubsec:MC} because of existence of two constraints given by Eqs.~(\ref{STeq:constraint_tsp_1}) and (\ref{STeq:constraint_tsp_2}).
The simplest way of transition between states is realized by flipping four spins simultaneously as shown in Fig.~\ref{STfig:TSP_protocol}.

Suppose we consider the case that we pass through at the $i$-th city at the $a$-th step and pass through at the $j$-th city at the $a'$-th step, which is described as
\begin{eqnarray}
 \sigma_{i,a}^z = +1,\,\,
  \sigma_{j,a}^z = -1,\,\,
  \sigma_{i,a'}^z = -1,\,\,
  \sigma_{j,a'}^z = +1.
\end{eqnarray}
The trial state generated by flipping four spins is as follows:
\begin{eqnarray}
 \sigma_{i,a}^z = -1,\,\,
  \sigma_{j,a}^z = +1,\,\,
  \sigma_{i,a'}^z = +1,\,\,
  \sigma_{j,a'}^z = -1.
\end{eqnarray}
The heat-bath method and the Metropolis method can be adopted for the transition probability between the present state and the trial state.
In Fig.~\ref{STfig:TSP_protocol}, $i=3$, $j=6$, $a=2$, and $a'=5$.

It should be noted that without loss of generality the initial condition can be set as
\begin{eqnarray}
  \sigma_{1,1} = +1,\,\,
  \sigma_{i,1} = -1 \qquad (i \neq 1),
\end{eqnarray}
and thus we can fix the states at the first step ($a=1$) during calculation.
The number of interactions in which we try to flip all spins in each Monte Carlo step is $(N-1)(N-2)/2$.

\subsubsection{Quantum Annealing}

In order to perform the quantum annealing, we introduce the transverse field as the quantum fluctuation effect as shown in Sec.~\ref{STsec:implementation}.
The quantum Hamiltonian is given by
\begin{eqnarray}
 \hat{\cal H} = \frac{1}{4} \sum_{a=1}^N \sum_{i,j} \ell_{i,j} \hat{\sigma}_{i,a}^z \hat{\sigma}_{j,a+1}^z
  -\Gamma \sum_{a=1}^N  \sum_{i=1}^N \hat{\sigma}_{i,a}^x,
\end{eqnarray}
where the first-term corresponds to the length of path and the second-term denotes the transverse field.
We can map this quantum Hamiltonian on $N\times N$ two-dimensional lattice onto $N\times N\times m$ three-dimensional Ising model as well as the case which was considered in Sec.~\ref{STsubsec:MC}.
The effective classical Hamiltonian derived by the Suzuki-Trotter decomposition is written as
\begin{align}
{\cal H}_{\rm eff} &= \frac{1}{4m} \sum_{a=1}^N \sum_{i,j} \sum_{k=1}^m \ell_{i,j} \sigma_{i,a,k}^z \sigma_{j,a+1,k}^z \notag \\
&- \frac{1}{\beta} \sum_{a=1}^N \sum_{i=1}^N \sum_{k=1}^m \frac{1}{2} \ln \coth \left( \frac{\beta \Gamma}{m} \right) \sigma_{i,a,k}^z \sigma_{i,a,k+1}^z, \ \ \ \ \sigma_{i,a,k}^z = \pm 1.
\end{align}
In the quantum annealing procedure, we have to take care of the constraints given by Eqs.~(\ref{STeq:constraint_tsp_1}) and (\ref{STeq:constraint_tsp_2}) as stated before.
Then the simplest way of changing state is to flip simultaneously four spins on the same layer ($m$ is fixed) along the Trotter axis.

\subsubsection{Comparison with Simulated Annealing and Quantum Annealing}

\begin{figure}[b]
 \begin{center}
 \psfig{file=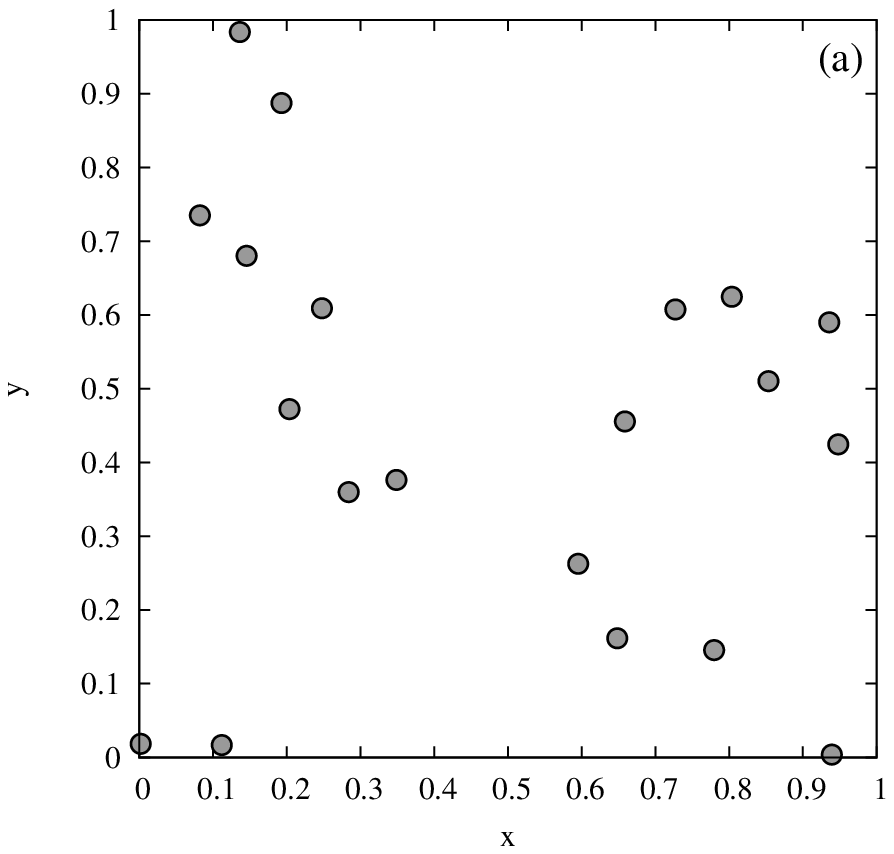,scale=0.59}
 \psfig{file=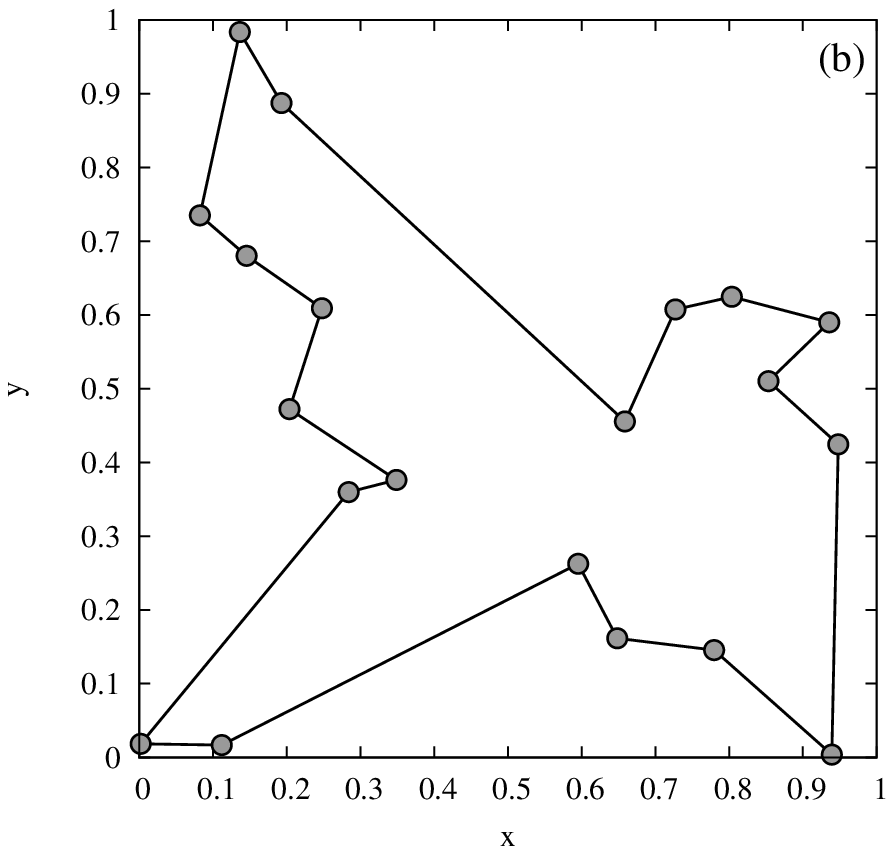,scale=0.59}
 \end{center}
 \caption{
 Traveling salesman problem for $N=20$.
 (a) Positions of cities.
 (b) The best solution in which the length of the path is minimum.
 }
 \label{STfig:tsp20}
\end{figure}

In order to demonstrate the comparison with the simulated annealing and the quantum annealing,
we perform the Monte Carlo simulation for the traveling salesman problem.
As an example,
we consider $N=20$ cities depicted in Fig.~\ref{STfig:tsp20} (a).
The positions of these cities were generated by pair of uniform random numbers ($0 \le x_i, y_i \le 1$).
The time schedules of temperature $T(t)$ for the simulated annealing and transverse field $\Gamma (t)$ for the quantum annealing are defined as
\begin{align}
T(t) &:= T_0 + T_1 \left( 1- \frac{t}{\tau} \right), \\
\Gamma (t) &:= \Gamma_0 + \Gamma_1 \left( 1- \frac{t}{\tau} \right),
\end{align}
where $T_0$ and $\Gamma_0$ are temperature and transverse field at the final time ($t=\tau$),
and $T_0+T_1$ and $\Gamma_0+\Gamma_1$ are temperature and transverse field at the initial time ($t=0$).
The value of $\tau^{-1}$ indicates the annealing speed,
and the annealing speed becomes slow as the value of $\tau$ increases.
In our simulations,
we adopt $T_0=\Gamma_0=0.01$ and $T_1=\Gamma_1=5$.
Furthermore, we fix the transverse field as $\Gamma=0$ during the simulation in the simulated annealing and the temperature as $T=0.01$ during the simulation in the quantum annealing.

\begin{figure}
 \begin{center}
  \psfig{file=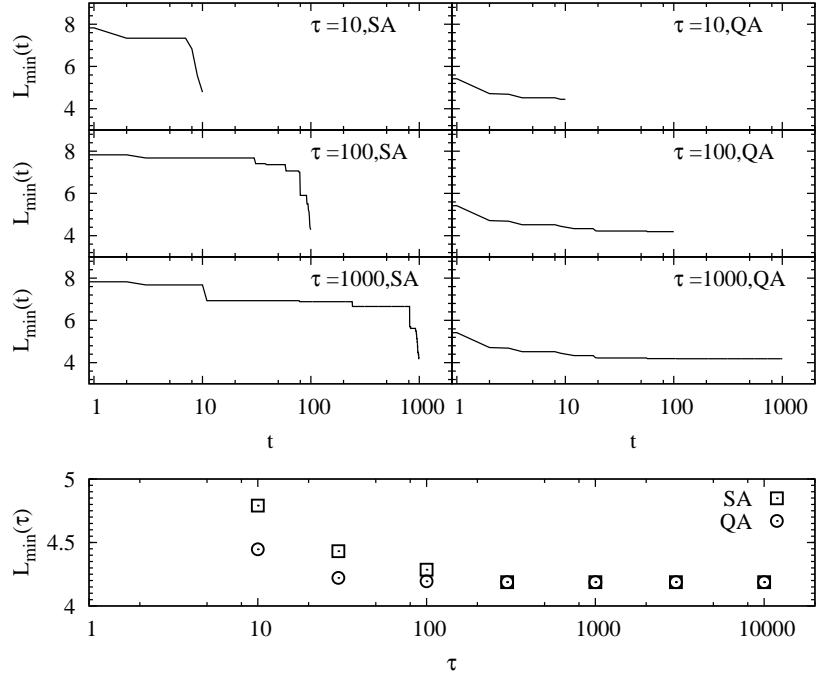,scale=0.95}
 \end{center}
 \caption{
 (Upper panel) Time dependence of minimum length of path $L_{\rm min} (t)$ for $\tau = 10$, $100$, and $1000$ obtained by the simulated annealing (SA) and the quantum annealing (QA).
 (Lower panel) Sweeping-time $\tau$ dependence of minimum length of path at the final state $L_{\rm min} (\tau)$ obtained by the simulated annealing indicated by squares and the quantum annealing indicated by circles.
 }
 \label{STfig:tsp20_time}
\end{figure}

We execute 100 independent simulations of simulated annealing based on the heat-bath type Monte Carlo method where each initial state generated by the uniform random number is different.
To compare the efficiency of the simulated annealing and quantum annealing in an equitable manner,
in the quantum annealing,
the Trotter number is putted as $m=10$, and we execute 10 independent simulations. 
We also calculate the minimum length of path $L_{\rm min} (t) := \min\{L(t')|0\le t' \le t\}$.
It should be noted that $L_{\rm min} (t)$ is a monotonic decreasing function.
The upper panel of Fig.~\ref{STfig:tsp20_time} shows the time dependence of minimum length of path $L_{\rm min} (t)$ for various $\tau$.
From the upper panel of Fig.~\ref{STfig:tsp20_time}, we can see that the convergence of minimum length of path in the quantum annealing is faster than that in the simulated annealing.
We also show the sweeping time $\tau$ dependence of the minimum length of path at the final state $L_{\rm min} (\tau)$ in the lower panel of Fig.~\ref{STfig:tsp20_time}.
This figure indicates that the obtained solution in the quantum annealing is always better than that in the simulated annealing.
Figure~\ref{STfig:tsp20} (b) shows the obtained best solution in both the simulated annealing and the quantum annealing with slow schedule.

In this way, we can obtain a better solution (in this case, the best solution) by both annealing methods with slow schedule.
Moreover, in our calculation,
the convergence of solution in the quantum annealing is faster than that in the simulated annealing,
and the obtained solution in the quantum annealing is better than that in the simulated annealing regardless of sweeping time $\tau$.
Thus, we can say that the quantum annealing method is appropriate as the annealing method for the traveling salesman problem
in comparison with the simulated annealing.
This fact has been confirmed in some researches\cite{STKadowaki-1998b,STMartonak-2004}.


\subsection{Clustering Problem}

In Sec.~\ref{STsec:tsp}, we explained the traveling salesman problem which can be mapped onto the Ising model with some constraints.
Many optimization problems can also be mapped onto the Ising model.
However, there are a number of optimization problems that can be described by the other models which are straightforward extensions of the Ising model.
In this section, we review the concept of clustering problem as such an example.

Clustering problem is also one of important optimization problems in information science and engineering\cite{STKurihara-2009,STSato-2009,STTanaka-2011a}.
We need to categorize much data in the real world according to its contents in various situations.
For instance, suppose we play stock market.
In order to see the socioeconomic situation, we want to extract efficiently important information related to stock market from an enormous quantity of information in news sites and newspapers.
In this case, it is better to categorize many articles in news sites and newspapers according to their contents.
This is an example of clustering problem which is adopted for many applications in wide area of science such as cognitive science, social science, and psychology.
The clustering problem is to divide the whole set into a couple of subsets.
Here we refer to the subsets as ``cluster''.

\begin{figure}[b]
 \begin{center}
  \psfig{file=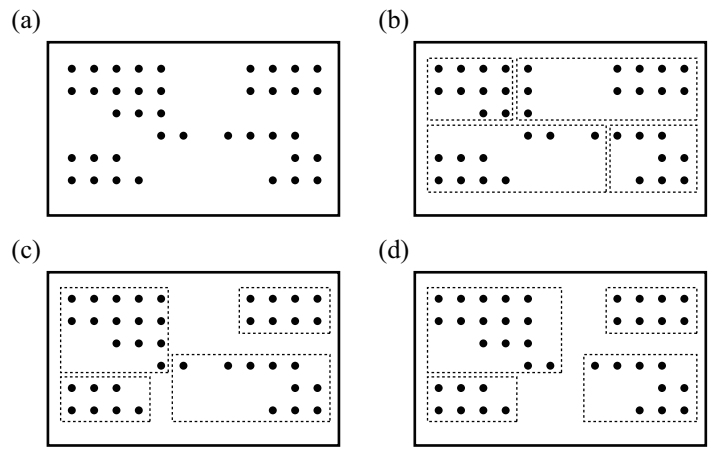,scale=1.0}
 \end{center}
 \caption{
 Schematic pictures of clustering problem.
 The points represent data and the square denote the whole set.
 (a) Data set.
 (b) Unstable solution $\Sigma_1$.
 (c) Metastable solution $\Sigma_2$.
 (d) The best solution $\Sigma_*$.
 }
 \label{STfig:schematicclustering}
\end{figure}

Figure~\ref{STfig:schematicclustering} shows schematic picture of the clustering problem.
Suppose we consider much data in the whole set which represents the square frame in Fig.~\ref{STfig:schematicclustering} (a).
The points in Fig.~\ref{STfig:schematicclustering} denote individual data.
In the clustering problem, our target is to find which the best division is.
Figure~\ref{STfig:schematicclustering} (b), (c), and (d) represent typical clustering states $\Sigma_1$, $\Sigma_2$, and $\Sigma_*$, respectively.
The states $\Sigma_1$ and $\Sigma_2$ are an unstable solution and a metastable solution, respectively.
The state $\Sigma_*$ denotes the best solution of clustering problem.

In order to consider how to implement the quantum annealing, the clustering problem can be described by the Potts model with random interactions\footnote{In practice, we do not know $\{J_{ij}\}$ and have to estimate interactions when we consider the clustering problem. However, we assume the Hamiltonian for simple explanation. As shown in this section, the implementation method does not depend on the specific form of interactions.}.
The Hamiltonian of the Potts model is given by
\begin{eqnarray}
 {\cal H}_{\rm Potts} = - \sum_{i,j} J_{ij} \delta_{\sigma_i,\sigma_j},
  \qquad
  \sigma_i = 1, \cdots, Q,
\end{eqnarray}
where the summation runs over all pairs of the $i$-th and $j$-th data.
The spin variable $\sigma_i$ represents individual data.
Here the value of $Q$ represents the number of clusters.
When $\sigma_i = \sigma_j$, the $i$-th and $j$-th data are in the same cluster.
It is natural to adopt ferromagnetic/antiferromagnetic interaction between data in the same/different cluster.
It should be noted that the Potts model is a straightforward extension of the Ising model since the Potts model is equivalent to the Ising model if $Q=2$.
Then the clustering problem is a problem to obtain the ground state of the Hamiltonian of the Potts model with given random interactions.
Here we assume that the number of clusters is fixed.

Next we explain how to introduce quantum field in order to perform the quantum annealing.
In optimization problems which can be represented by the Ising model, we can use transverse field as the quantum fluctuation which is represented as $-\Gamma \sum_i \sigma_i^x$.
However, we cannot use this transverse field $-\Gamma \sum_i \sigma_i^x$ for the clustering problem directly,
since the matrix which represents the state is $Q \times Q$ matrix.
Thus, we generalize the $x$-component of the Pauli matrix of the Ising model as follows:
\begin{eqnarray}
 \hat{\tau}^x := 
  \mathbb{E}_Q - \mathbb{I}_Q
  =
  \left(
   \begin{array}{ccccc}
   0 &-1 &-1 &\cdots & -1\\
   -1 &0 &-1 &\cdots & -1\\
   -1 &-1 &0 &\vdots &-1 \\
   \vdots & \vdots & \vdots & \ddots & \vdots\\
   -1 &-1 &-1 &\cdots & 0 
   \end{array}
  \right),
\end{eqnarray}
where $\mathbb{E}_Q$ and $\mathbb{I}_Q$ represent the $Q \times Q$ unit matrix and the $Q\times Q$ matrix whose all elements are unity.
By using this generalized Pauli matrix, we can apply the quantum annealing for clustering problem\cite{STKurihara-2009,STSato-2009,STTanaka-2011a}.
Here we consider the following Hamiltonian:
\begin{eqnarray}
 \hat{\cal H} = \hat{\cal H}_{\rm Potts} + \hat{\cal H}_{\rm q}^{\rm (Potts)},
  \qquad
 \hat{\cal H}_{\rm q}^{\rm (Potts)} := - \Gamma \sum_{i=1}^N \hat{\tau}_i^x,
\end{eqnarray}
where $N$ is the number of individual data.
As well as the case for the Ising model, we can calculate the partition function of the Hamiltonian:
\begin{eqnarray}
 \nonumber
 Z_{\rm Potts} &&= {\rm Tr}\, {\rm e}^{-\beta\hat{\cal H}} = \sum_{\Sigma}
  \Braket{\Sigma|{\rm e}^{-\beta(\hat{\cal H}_{\rm Potts}+\hat{\cal H}_{\rm q}^{\rm (Potts)})}|\Sigma}\\
 \nonumber
 &&= \lim_{m\to \infty}\sum_{\{\Sigma_k\},\{\Sigma'_k\}}
  \Braket{\Sigma_1|{\rm e}^{-\beta\hat{\cal H}_{\rm Potts}/m}|\Sigma_1'}
  \Braket{\Sigma_1'|{\rm e}^{-\beta\hat{\cal H}_{\rm q}^{\rm (Potts)}/m}|\Sigma_2}\\
 \nonumber  
  &&\times \Braket{\Sigma_2|{\rm e}^{-\beta\hat{\cal H}_{\rm Potts}/m}|\Sigma_2'}
  \Braket{\Sigma_2'|{\rm e}^{-\beta\hat{\cal H}_{\rm q}^{\rm (Potts)}/m}|\Sigma_3}\\
 &&\times 
  \Braket{\Sigma_m|{\rm e}^{-\beta\hat{\cal H}_{\rm Potts}/m}|\Sigma_m'}
  \Braket{\Sigma_m'|{\rm e}^{-\beta\hat{\cal H}_{\rm q}^{\rm (Potts)}/m}|\Sigma_1},
\end{eqnarray}
where $\ket{\Sigma_k}$ represents the direct-product space of $N$ spins:
\begin{eqnarray}
 \ket{\Sigma_k} = \ket{\sigma_{1,k}} \otimes \ket{\sigma_{2,k}} \otimes \cdots \ket{\sigma_{N,k}}.
\end{eqnarray}
There are two elements $\braket{\Sigma_k|{\rm e}^{-\beta\hat{\cal H}_{\rm Potts}/m}|\Sigma_k'}$ and $\braket{\Sigma_{k}'|{\rm e}^{-\beta\hat{\cal H}_{\rm q}^{\rm (Potts)}/m}|\Sigma_{k+1}}$.
These factors are calculated as follows:
\begin{eqnarray}
 &&\Braket{\Sigma_k|{\rm e}^{-\beta\hat{\cal H}_{\rm Potts}/m}|\Sigma_k'}
  = \exp
  \left(
   \frac{\beta}{m} \sum_{i,j} J_{ij} \delta_{\sigma_{i,k},\sigma_{j,k}} 
  \right)
  \prod_{i=1}^N \delta_{\sigma_{i,k},\sigma_{i,k}'},\\
 \nonumber 
  &&\Braket{\Sigma_{k}'|{\rm e}^{-\beta\hat{\cal H}_{\rm q}^{\rm (Potts)}/m}|\Sigma_{k+1}}
  = \prod_{i=1}^N 
  \left[
   {\rm e}^{-\frac{\beta\Gamma}{m}} \delta_{\sigma_{i,k}'\sigma_{i,k+1}} + 
   \frac{1}{Q} \left( {\rm e}^{-\frac{\beta \Gamma}{m}(1-Q)} - 1\right)
  \right].\\
 \,
\end{eqnarray}
By using the above expressions, we can perform the quantum Monte Carlo simulation as well as the Ising model with transverse field.
If the spin variable is not $S=1/2$ Ising spin as in the case just described, we can implement the quantum annealing by considering appropriate quantum field.
There are some studies that the quantum annealing succeeds to obtain the better solution than the simulated annealing for clustering problems\cite{STKurihara-2009,STSato-2009,STTanaka-2011a}.
\section{Relationship between Quantum Annealing and Statistical Physics}
\label{STsec:QA_SM}

In the preceding sections we explained the Ising model, a couple of implementation methods of the quantum annealing, and the optimization problems.
There are a couple of studies that clarify the efficiency and feature of the quantum annealing in terms of statistical physics.
In this section we take two examples which display relationship between quantum annealing and statistical physics focusing on the thermal fluctuation effect and the quantum fluctuation effect for ordering phenomena.
In the first half, we review the Kibble-Zurek mechanism which characterizes the efficiency of the quantum annealing for systems where a second-order phase transition occurs comparing with the efficiency of the simulated annealing.
In the last half, we show similarities and differences between thermal fluctuation and quantum fluctuation for frustrated Ising spin systems.

\subsection{Kibble-Zurek Mechanism}
\label{STsec:KZmechanism}

In statistical physics, it has been an important topic to investigate the ordering process in systems where a phase transition takes place\cite{STTakayama-2007,STMiyashita-2007,STTanaka-2007,STTanaka-2007b,STTanaka-2009,STTanaka-2010,STTanaka-2011b}.
Especially, dynamical properties during changing control variables such as temperature and external fields are interesting\cite{STTakayama-2007,STMiyashita-2007,STTanaka-2010}.
Recently, the Kibble-Zurek mechanism has been drawing attention not only in statistical physics and condensed matter physics but also for the quantum annealing.
In this section,
we explain the Kibble-Zurek mechanism relating to a dynamics which passes across a second-order phase transition point.
The Kibble-Zurek mechanism can make clear what happens in systems where the second-order phase transition occurs during the simulated annealing and the quantum annealing from a viewpoint of statistical physics.
Before we consider the efficiency of the quantum annealing comparing with the simulated annealing by using the Kibble-Zurek mechanism, we show the general feature of the Kibble-Zurek mechanism.

As an example,
we consider the Kibble-Zurek mechanism in the ferromagnetic system where the second-order phase transition occurs at finite temperature.
At the second-order phase transition point,
the correlation length diverges in the equilibrium state,
and thus the relaxation time should be infinite.
Hence,
the system cannot reach the equilibrium state,
when we decrease temperature to the transition temperature with finite speed.
Furthermore,
since the relaxation time is long around the transition temperature,
it is difficult to equilibrate the system.
Here,
we assume that growth of correlation length stops at the temperature where the system is less able to reach the equilibrium state.
If we decrease temperature slow enough,
the system can reach the equilibrium state even near the transition point.
Thus,
it is expected that the value of stopped correlation length because of the long relaxation time depends on the annealing speed.
As we will see below, the value of stopped correlation length can be scaled by the annealing speed.

To consider the second-order phase transition at finite temperature in the ferromagnetic systems,
we define the dimensionless temperature $g$ as
\begin{align}
g := \frac{T-T_\text{c}}{T_\text{c}},
\end{align}
where $T_\text{c}$ is the phase transition temperature.
When the absolute value of $g$ is small, it is believed that the scaling ansatz is valid.
By the scaling ansatz,
the temperature-dependent correlation length $\xi (g)$ is given as\cite{STNishimori-2011}
\begin{align}
\xi(g) \propto |g|^{-\nu}, \label{STkibble:eq:xi}
\end{align}
where $\nu$ is one of the critical exponents.
Moreover,
the relaxation time $\tau_\text{rel}$ is scaled by the following relation\cite{STNishimori-2011}:
\begin{align}
\tau_\text{rel} (g) \propto [\xi(g)]^z \propto |g|^{-z\nu}, \label{STkibble:eq:tau}
\end{align}
where $z$ is the dynamical critical exponent.
Here,
we decrease the temperature $T(t)$ against the time $t$ as following schedule:
\begin{align}
T(t) = T_\text{c} \left( 1 - \frac{t}{\tau_\text{Q}} \right) \ \ \ \ \ \ (-\infty < t \le \tau_\text{Q}). \label{STkibble:eq:time_dep}
\end{align}
The value of $\tau_\text{Q}^{-1}$ corresponds to the annealing speed.
When the value of $\tau_\text{Q}$ is large/small,
the system is annealed to low temperature slowly/quickly.
At $t=0$,
the temperature is the phase transition temperature ($T(0)=T_\text{c}$),
and the temperature is zero ($T(\tau_\text{Q})=0$) at $t=\tau_\text{Q}$.
From Eq.~(\ref{STkibble:eq:time_dep}),
the dimensionless temperature $g$ becomes the time-dependent function as follows:
\begin{align}
g(t) = \frac{T(t)-T_\text{c}}{T_\text{c}} = - \frac{t}{\tau_\text{Q}}.
\end{align}

\begin{figure}[b]
 \begin{center}
  \psfig{file=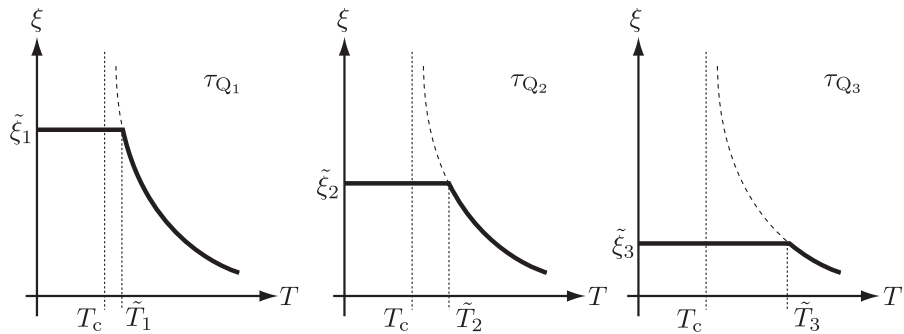,scale=0.85}
  \end{center}
 \caption{
 Schematic of the annealing speed dependence of correlation length $\xi (g(t))$.
 $\tau_\text{Q}^{-1}$ is annealing speed and $\tau_{\text{Q}_1} > \tau_{\text{Q}_2} > \tau_{\text{Q}_3}$.
 We define $\tilde{T}_i:=T_{\rm c}(1+|\tilde{t}|/\tau_{{\rm Q}_i})$ and $\tilde{\xi}_i:=\xi(|\tilde{t}|/\tau_{{\rm Q}_i})$.
 The dotted curve represents correlation length in the equilibrium state.
 }
 \label{STfig:Kibble-correlation}
\end{figure}

In the Kibble-Zurek mechanism,
we assume the following situation:
\begin{align}
\begin{cases}
\tau_\text{rel} (g(t)) < |t| & : \text{system {\it can} reach equilibrium state} \\
\tau_\text{rel} (g(t)) > |t| & : \text{system {\it cannot} reach equilibrium state} 
\end{cases},
\end{align}
where $|t|$ is a remaining time to transition temperature.
That is,
when a remaining time $|t|$ is longer/shorter than the relaxation time $\tau_\text{rel} (g(t))$,
the system can/cannot reach the equilibrium state.
Note that the value of considered $t$ should be negative since the relaxation time diverges before the temperature reaches the transition temperature ($t=0$).
From this assumption,
the time $\tilde{t}$ at which the system is less able to reach the equilibrium state is defined by following relation:
\begin{align}
\tau_\text{rel} (g(\tilde{t})) = |\tilde{t}|.
\end{align}
Furthermore, since we have assumed that the growth of correlation length stops at $t=\tilde{t}$,
the value of correlation length is always $\xi (g(\tilde{t}))$ below $T(\tilde{t})$ as shown in Fig.~\ref{STfig:Kibble-correlation}.
Moreover,
the dimensionless temperature at $\tilde{t}$ is expressed as
\begin{align}
g(\tilde{t}) = \frac{|\tilde{t}|}{\tau_\text{Q}} = \frac{\tau_\text{rel} (g(\tilde{t}))}{\tau_\text{Q}} \propto \frac{|g (\tilde{t})|^{-z\nu}}{\tau_\text{Q}}. \label{STkibble:eq:kibble_temp}
\end{align}
From this relation,
$g(\tilde{t})$ is scaled by the annealing speed, and from Eqs.~(\ref{STkibble:eq:xi}) and (\ref{STkibble:eq:kibble_temp}),
 the correlation length at $t=\tilde{t}$ is scaled as follows:
\begin{align}
 g(\tilde{t}) \propto \tau_\text{Q}^{-\frac{1}{1+z\nu}},
 \qquad
 \xi(g(\tilde{t})) \propto \tau_\text{Q}^{\frac{\nu}{1+z\nu}}.
\end{align}
Furthermore,
the density of domain wall $n(t)$ is written as
\begin{align}
n(t) \propto \xi(g(t))^{-d},
\end{align}
where $d$ is the spatial dimension,
and $n(\tilde{t})$ at $t=\tilde{t}$ is scaled as follows:
\begin{align}
\label{STeq:KZrelation_exponents}
n(\tilde{t}) \propto \tau_\text{Q}^{-\frac{d\nu}{1+z\nu}}.
\end{align}
For instance,
in the ferromagnetic Ising model on two-dimensional lattice ($d=2$, $\nu=1$) when we adopt the Monte Carlo dynamics based on the single-spin-flip method ($z=2.132$)\cite{STIto-1987},
the correlation length and the density of domain wall at $t=\tilde{t}$ are naively obtained as
\begin{align}
 \xi(g(\tilde{t})) \propto \tau_\text{Q}^{0.319},
 \qquad
 n(\tilde{t}) \propto \tau_\text{Q}^{-0.639}.
\end{align}

In this way,
in the dynamics which passes across the second-order phase transition point at finite temperature,
the correlation length and the density of domain wall (topological defect) are scaled by the annealing speed.
This argument is called the Kibble-Zurek mechanism.
Since the Kibble-Zurek mechanism explains the creation of topological defects induced by cooling of the system which takes place the second-order phase transition,
this relates to the evolution of cosmic strings by spontaneous symmetry breaking in the Big Bang theory\cite{STKibble-1976,STKibble-1980,STZurek-1985}.
The Kibble-Zurek mechanism can also describe the creation of topological defects in magnetic models\cite{STDamski-2005,STZurek-2005}, superfluid helium systems\cite{STRuutu-1996,STEltsov-2000}, and Bose-Einstein condensations\cite{STSaito-2007,STWeiler-2008}.
Next we consider the efficiency of the simulated annealing and the quantum annealing using the Kibble-Zurek mechanism by taking examples which can be treated analytically.

\subsubsection{Efficiency of Simulated Annealing and Quantum Annealing}

Next, we consider the efficiency of the simulated annealing and the quantum annealing according to the Kibble-Zurek mechanism.
As an example,
we treat the case where the non-domain wall state is the best solution.
In this case,
the value of $n(\tilde{t})$ approximately represents the difference between the obtained solution and the best solution.
Thus, by using the Kibble-Zurek mechanism, we can compare the efficiency of annealing methods from the behavior of $n(\tilde{t})$ against the annealing speed.
Suppose we solve optimization problems by using annealing methods, we would like to obtain a better solution as fast as possible, in other words, as small $\tau_{\rm Q}$ as possible.
Then, the comparison obtained by the Kibble-Zurek mechanism is expected to become an useful information for the optimization problems.

%
%
As an example,
we consider the efficiency of the simulated annealing and the quantum annealing for the random ferromagnetic Ising chain in terms of the Kibble-Zurek mechanism according to Refs.~[\refcite{STSuzuki-2009,STSuzuki-2011a}].

\subsubsection{Simulated Annealing for Random Ferromagnetic Ising Chain}

The model Hamiltonian of the random ferromagnetic Ising chain is given as
\begin{align}
\mathcal{H} = - \sum_i J_i \sigma_i^z \sigma_{i+1}^z, \ \ \ \sigma_i^z = \pm 1,
\end{align}
where $J_i$ is the interaction between the $i$-th site and the $(i+1)$-th site.
The value of $J_i$ is given by the uniform distribution between $0 < J_i \le 1$.
The distribution function $P^{\rm (u)} (J_i)$ is given by
\begin{align}
P^{\rm (u)} (J_i) := \begin{cases} 1 & \text{for} \ 0 < J_i \le 1 \\ 0 & \text{otherwise} \end{cases}.
\end{align}
Since the interaction $J_i$ is always positive value,
the ground state spin configuration is the all-up spin state or the all-down spin state.
In this model, the ferromagnetic transition occurs at zero temperature.

The correlation function between two sites where the distance is $r$ is written as
\begin{align}
[\langle \sigma_i \sigma_{i+r} \rangle ]_{\text{av}} = \left( \frac{1}{\beta} \ln \cosh \beta \right)^r,
\end{align}
where $\langle \cdots \rangle$ and $[\cdots]_{\rm av}$ denote the thermal average and the random average.
Physical quantities should depend on the specific spatial pattern of the random interactions $\{J_i\}$.
Then, these averages are defined by
\begin{align}
 \langle O(\{J_i\}) \rangle &:=
  \frac{{\rm Tr}\, O(\{J_i\}) {\rm e}^{-\beta \mathcal{H}}}{{\rm Tr}\, {\rm e}^{-\beta \mathcal{H}}}, \\
 [O(\{J_i\})]_\text{av} &:= \int \prod_{i} {\rm d} J_i P^{\rm (u)}(J_i) O(\{J_i\}),
\end{align}
respectively.
We omit the argument $(\{J_i\})$ for simplicity.
The relationship between the correlation function and the correlation length $\xi$ is given by
\begin{align}
[\langle \sigma_i \sigma_{i+r} \rangle ]_{\text{av}} = {\rm e}^{-r/\xi}.
\end{align}
Here we mainly focus on the low-temperature limit,
since the correlation length grows as temperature decreases.
Then the correlation length is given as
\begin{align}
\xi = - \frac{1}{\ln (\beta^{-1} \ln \cosh \beta)} \simeq \frac{\beta}{\ln 2}. \label{STkibble:eq:sa_xi}
\end{align}
Here,
we adopt the Glauber dynamics\cite{STGlauber-1963} as the time development,
and thus the relaxation time $\tau_\text{rel}$ can be written as
\begin{align}
\tau_\text{rel}= \frac{1}{1-\tanh 2 \beta} \simeq \frac{1}{2} {\rm e}^{4 \beta} = \frac{1}{2} {\rm e}^{4 \xi \ln 2}. \label{STkibble:eq:sa_tau}
\end{align}
As we can see, in this model,
the correlation length $\xi$ and the relaxation time $\tau_{\rm rel}$ are not the power function of temperature unlike the case of the systems where the second-order phase transition occurs at finite temperature (Eqs.~(\ref{STkibble:eq:xi}) and (\ref{STkibble:eq:tau})).
This is because properties are different between phase transition which exhibits at finite temperature and that occurs at zero temperature.

We decrease temperature $T(t)$ against the time $t$ as following schedule:
\begin{align}
T(t) = - \frac{t}{\tau_\text{Q}} \ \ \ \ \ \ (-\infty < t \le 0).
\end{align}
Here $T_\text{c}=0$ in this system.
According to the Kibble-Zurek mechanism,
we define $\tilde{t}$ by following relation:
\begin{align}
\tau_\text{rel} (T(\tilde{t})) = |\tilde{t}|,
\end{align}
and, we obtain
\begin{align}
T(\tilde{t}) = \frac{|\tilde{t}|}{\tau_\text{Q}} = \frac{\tau_\text{rel} (T(\tilde{t}))}{\tau_\text{Q}}. \label{STkibble:eq:sa_ttilde}
\end{align}
By using Eqs.~(\ref{STkibble:eq:sa_xi}) and (\ref{STkibble:eq:sa_tau}),
low-temperature limit of Eq.~(\ref{STkibble:eq:sa_ttilde}) is written as
\begin{align}
\frac{1}{\xi (T(\tilde{t}))\ln 2} \simeq \frac{1}{2 \tau_\text{Q}} {\rm e}^{4 \xi (T(\tilde{t}))\ln 2},
\end{align}
and, we obtain
\begin{align}
\xi (T(\tilde{t})) &= \frac{\ln \tau_\text{Q} + \ln 2 - \ln (\xi (T(\tilde{t})) \ln 2)}{4 \ln 2} \propto \frac{\ln \tau_\text{Q}}{4 \ln 2}.
\end{align}
The approximation of RHS is valid in the case of $\tau_\text{Q} \gg 1$ which indicates very slow annealing speed.
Thus,
we can estimate the density of domain wall $n_\text{SA}(\tilde{t})$ at $t=\tilde{t}$ as follows:
\begin{align}
n_\text{SA}(\tilde{t}) \propto \frac{4 \ln 2}{\ln \tau_\text{Q}}.
\end{align}

\subsubsection{Quantum Annealing for Random Ferromagnetic Ising Chain}

We study the Kibble-Zurek mechanism for the random ferromagnetic Ising chain with transverse field $\Gamma$.
The model Hamiltonian is given as
\begin{align}
\hat{\cal{H}} = - \sum_i J_i \hat{\sigma}_i^z \hat{\sigma}_{i+1}^z - \Gamma \sum_i \hat{\sigma}_i^x ,
\end{align}
where the value of $J_i$ is given by the uniform distribution between $0 < J_i \le 1$ as well as the case of simulated annealing.
In this model,
the quantum phase transition from the paramagnetic phase to the ferromagnetic phase occurs at $\Gamma_{\rm c} = \exp ([\ln J_i]_\text{av})$\cite{STShankar-1987}.
Here,
we define the dimensionless transverse field $g$ as 
\begin{align}
g := \frac{\Gamma-\Gamma_\text{c}}{\Gamma_\text{c}}.
\end{align}
When $|g| \ll 1$,
it has been known that the correlation length obtained by the renormalization group analysis\cite{STFisher-1995} is scaled by the following relation:
\begin{align}
\xi (g) \propto |g|^{-\nu} \ \ \ \ \ (\nu=2). \label{STkibble:eq:qa_xi}
\end{align}
Moreover,
a coherence time $\tau_\text{coh}$ is scaled by
\begin{align}
\tau_\text{coh} (g) \propto [\xi (g)]^z \propto |g|^{-\nu z} \ \ \ \ \ (\nu=2), \label{STkibble:eq:qa_tau}
\end{align}
where the dynamical exponent $z$ is scaled as
\begin{align}
z \propto \frac{1}{|g|}, \label{STkibble:eq:qa_z}
\end{align}
which is also obtained by the renormalization group analysis\cite{STFisher-1995}.
This means that the dynamical exponent diverges at the transition point,
and this behavior is a qualitative difference between the random system and the pure system ($z=1$).
From this fact, $\tau_{\rm coh}$ cannot be expressed by the power function of $g$ unlike the case of the second-order phase transition at finite temperature.

We decrease transverse field $\Gamma (t)$ against the time $t$ as following schedule:
\begin{align}
\Gamma (t) = \Gamma_\text{c} \left( 1 - \frac{t}{\tau_\text{Q}} \right) \ \ \ \ \ \ (-\infty < t \le \tau_\text{Q}).
\end{align}
According to the Kibble-Zurek mechanism,
we define $\tilde{t}$ by following relation:
\begin{align}
\tau_\text{coh} (g(\tilde{t})) = |\tilde{t}|,
\end{align}
and we obtain
\begin{align}
g(\tilde{t}) = \frac{|\tilde{t}|}{\tau_\text{Q}} = \frac{\tau_\text{coh} (g(\tilde{t}))}{\tau_\text{Q}}. \label{STkibble:eq:qa_gtilde}
\end{align}
By using Eqs.~(\ref{STkibble:eq:qa_xi}), (\ref{STkibble:eq:qa_tau}), and (\ref{STkibble:eq:qa_z}),
Eq.~(\ref{STkibble:eq:qa_gtilde}) is written as
\begin{align}
\frac{1}{\sqrt{\xi (g(\tilde{t}))}} &\propto \frac{1}{\tau_\text{Q}} |\xi (g(\tilde{t}))|^z \propto \frac{1}{\tau_\text{Q}} |\xi (g(\tilde{t}))|^{\sqrt{\xi (g(\tilde{t}))}},
\end{align}
and, we obtain
\begin{align}
\left( \sqrt{\xi (g(\tilde{t}))} + \frac{1}{2} \right) \ln \xi (g(\tilde{t})) \propto \ln \tau_\text{Q}.
\end{align}
In the limit of $\tau_{\rm Q} \gg 1$,
since the value of $\xi (g(\tilde{t}))$ is very large,
\begin{align}
\sqrt{\xi (g(\tilde{t}))} + \frac{1}{2} \simeq \sqrt{\xi (g(\tilde{t}))},
\end{align}
and we obtain\cite{STDziarmaga-2006}
\begin{align}
\xi (g(\tilde{t})) \propto \left( \frac{\ln \tau_\text{Q}}{\ln \xi (g(\tilde{t}))} \right)^2.
\end{align}
Moreover,
since the change of $\ln \xi (g(\tilde{t}))$ is gradual in comparison with that of $\xi (g(\tilde{t}))$,
we neglect $\ln \xi (g(\tilde{t}))$ and obtain
\begin{align}
\xi (g(\tilde{t})) \propto \left( \ln \tau_\text{Q} \right)^2.
\end{align}
From this relation,
we can estimate the density of domain wall $n_\text{QA}(\tilde{t})$ at $t=\tilde{t}$ as follows:
\begin{align}
n_\text{QA}(\tilde{t}) \propto \left( \ln \tau_\text{Q} \right)^{-2}.
\end{align}

\subsubsection{Comparison between Simulated and Quantum Annealing Methods}

We have shown analysis of the domain wall density in the random ferromagnetic Ising chain during the simulated annealing and the quantum annealing by the Kibble-Zurek mechanism.
The obtained densities of domain wall are
\begin{align}
n_\text{SA}(\tilde{t}) &\propto \left( \ln \tau_\text{Q} \right)^{-1} \ \ : \ \ \text{simulated annealing}, \\
n_\text{QA}(\tilde{t}) &\propto \left( \ln \tau_\text{Q} \right)^{-2} \ \ : \ \ \text{quantum annealing}.
\end{align}
From these relations,
it is clear that the decay of $n_\text{QA}(\tilde{t})$ is faster than that of $n_\text{SA}(\tilde{t})$ against the value of $\tau_\text{Q}$.
Thus,
from the Kibble-Zurek mechanism,
it is concluded that the quantum annealing method is appropriate as the annealing method for the random ferromagnetic Ising chain in comparison with the simulated annealing method.
Suppose we consider the ferromagnetic Ising chain with homogeneous interaction ($J_i=1$ for all $i$).
In this case, both the domain wall density in the simulated annealing and that in the quantum annealing are obtained as
\begin{equation}
 n(\tilde{t}) \propto \frac{1}{\sqrt{\tau_{\rm Q}}}.
\end{equation}
This relation for the simulated annealing can be obtained by a simple calculation as well as the case of the random Ising spin chain.
On top of that, the relation for the quantum annealing can be derived by Eq.~(\ref{STeq:KZrelation_exponents}).
Here the critical exponent $\nu$ of the transverse Ising chain with homogeneous interaction is $\nu=1$ and the dynamical exponent of this system is $z=1$.
Then there is no difference between the simulated annealing and the quantum annealing in the case of the homogeneous ferromagnetic Ising chain.
However, since the optimization problem has some kind of randomness, the abovementioned result encourages that the quantum annealing is better than the simulated annealing for optimization problems. 


In general,
the existence of the phase transition in optimization problems negatively influences performance of annealing methods.
Here,
we have introduced the Kibble-Zurek mechanism relating to the dynamics which passes across the second-order phase transition point.
As the specific example,
we have analyzed the efficiencies of the simulated annealing and the quantum annealing for the random ferromagnetic Ising chain according to the Kibble-Zurek mechanism.
For this model,
the efficiency of the quantum annealing is better than that of the simulated annealing.
Of course,
since the efficiency of annealing methods depends on the details of optimization problems,
it is not to say that the quantum annealing is always appropriate as the annealing method for general optimization problems in comparison with the simulated annealing.
Moreover, we have to develop a theory based on the Kibble-Zurek mechanism itself\cite{STBiroli-2010}, since we assume the growth of the correlation length stops at $t>\tilde{t}$.
For example, if we adapt the Kibble-Zurek mechanism to two- or three-dimensional models and more complicated models,
it is difficult to estimate the correlation length analytically,
and thus we should execute numerical simulations such as the Monte Carlo simulation.
For example,
in the two-dimensional Ising model with random interactions,
it has been shown that the efficiency of the quantum annealing is better than that of the simulated annealing by Monte Carlo simulation.\cite{STSuzuki-2011a}
Although the efficiency of annealing methods for a number of optimization problems has been clarified by the Kibble-Zurek mechanism, it remains to be an open problem to investigate when to use the quantum annealing exhaustively.

In the above-mentioned argument,
the phase transition under consideration is of the second order.
What happens if we adapt the same argument for the other type phase transitions such as first-order phase transition and Kosterlitz-Thouless (KT) transition?
In these phase transitions,
the behaviors of correlation length are different from that in systems where a second-order phase transition occurs:
the finite-correlation length at the first-order phase transition point and the quasi-long-range correlation length at the KT transition point.
Thus,
it is an interesting problem to clarify relationship between behaviors of correlation length and the {\it generalized} Kibble-Zurek mechanism.
By considering dynamical nature of the optimization problems in terms of non-equilibrium statistical physics in a deeper way, we believe that the quantum annealing method will become a central part of practical method for optimization problems.


\subsection{Frustration Effects for Simulated Annealing and Quantum Annealing}
\label{STsec:Frustration}

\begin{figure}[b]
 \begin{center}
 \psfig{file=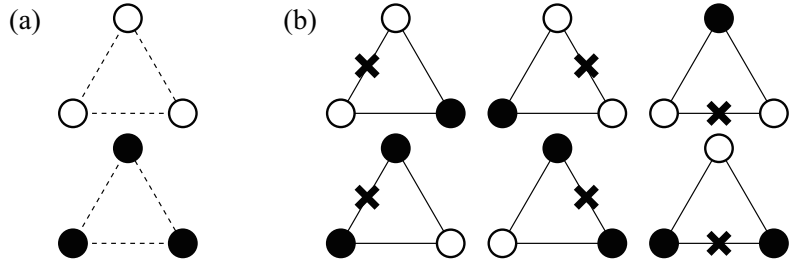, scale=0.9}
 \end{center}
 \caption{
 Three spin system on triangle cluster.
 The dotted and solid lines represent ferromagnetic and antiferromagnetic interactions, respectively.
 The open and solid circles are the $+1$-state and the $-1$-state, respectively.
 The crosses indicate the positions of unfavorable interactions.
 (a) Ground states for ferromagnetic case.
 (b) Ground states for antiferromagnetic case.
 }
 \label{STfig:trianglecluster}
\end{figure}

In many cases optimization problems can be represented by the Ising model with random interactions and magnetic fields as mentioned before.
The Hamiltonian of this system is given by
\begin{eqnarray}
 {\cal H} = - \sum_{i,j} J_{ij} \sigma_i^z \sigma_j^z - \sum_{i=1}^N h_i \sigma_i^z, \, \ \ \ \sigma_i^z = \pm 1.
\end{eqnarray}
When all interactions are ferromagnetic as the previous example in Sec.~\ref{STsec:KZmechanism}, the ground state is the all-up or the all-down states.
However, if there are antiferromagnetic interactions in the system, the situation becomes different.
In order to show the difference between ferromagnetic interaction and antiferromagnetic interaction, we first consider three spin system on triangle cluster as shown in Fig.~\ref{STfig:trianglecluster}.
In this section, we treat the case for $h_i=0$ for all $i$.
The dotted and solid lines in Fig.~\ref{STfig:trianglecluster} represent ferromagnetic and antiferromagnetic interactions, respectively.

The considered Hamiltonian is written as
\begin{eqnarray}
 {\cal H}_{\rm triangle} = -J (\sigma_1^z \sigma_2^z + \sigma_2^z \sigma_3^z + \sigma_3^z \sigma_1^z).
\end{eqnarray}
Here we set the all interactions are the same value for simplicity.
The ground states for positive $J$ (ferromagnetic interaction) are the all-up or the all-down states shown in Fig.~\ref{STfig:trianglecluster} (a).
In these states, all spins between all interactions are energetically favorable states.
In the case of negative $J$ (antiferromagnetic interaction), while on the other hand, six states shown in Fig.~\ref{STfig:trianglecluster} (b) are ground states.
These ground states have unfavorable interactions indicated by the crosses in Fig.~\ref{STfig:trianglecluster} (b).
This situation is called frustration.
In the homogeneous antiferromagnetic Ising spin systems on lattices based on triangle such as triangular lattice and kagom\'e lattice, frustration appears in all triangles.
Since such frustration comes from lattice geometry, this is called geometrical frustration.
It should be noted that the homogeneous antiferromagnetic Ising spin systems on square lattice and hexagonal lattice have no frustration.
Since these systems are bipartite systems which can be decomposed by two sublattices, these systems can be transformed on the ferromagnetic systems by local gauge transformation of all spins belonging to one of the sublattices.

\begin{figure}[b]
 \begin{center}
 \psfig{file=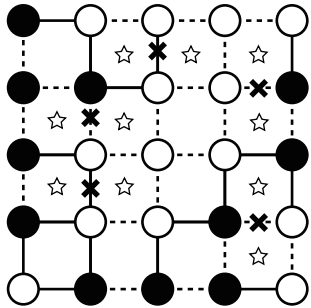, scale=1.0}
 \end{center}
 \caption{
 A ground state of the Ising spin system with random interactions.
 The dotted and solid lines represent ferromagnetic and antiferromagnetic interactions, respectively.
 The open and solid circles are the $+1$-state and the $-1$-state, respectively.
 The stars and crosses indicate frustration plaquettes and unfavorable interactions, respectively.
 }
 \label{STfig:randomfrustration}
\end{figure}

Frustration appears in also inhomogeneous systems as shown in Fig.~\ref{STfig:randomfrustration}.
The squares pointed by stars in Fig.~\ref{STfig:randomfrustration} represent frustration plaquettes which are satisfied following relation:
\begin{eqnarray}
 \kappa_k := \prod_{i,j \in \square_k} J_{ij} < 0,
\end{eqnarray}
where $\square_k$ indicates the smallest square plaquette at the position $k$.
If $\kappa_k$ for all $k$ is positive, the system is not frustrated.

In general, frustration prevents the system from conventional magnetic ordering such as ferromagnetic order and N\'eel order, since there is no state where all interactions are satisfied energetically in frustrated systems.
Frustration makes peculiar density of states which induces unconventional phase transition and slow dynamics\cite{STToulouse-1977,STLiebmann-1986,STKawamura-1998,STDiep-2005,STTanaka-2005,STTanaka-2007b,STTanaka-2007c,STTamura-2008,STTanaka-2009b,STTanaka-2010,STTamura-2011,STTamura-2011b}.
Although many optimization problems can be represented by the Ising model with random interactions and magnetic fields, here we focus on the frustration effect which comes from non-random interactions.
In terms of statistical physics, this is a first-step study to investigate similarities and differences between thermal fluctuation and quantum fluctuation for frustrated systems.
Furthermore, it is important topic for the optimization problems to consider the thermal fluctuation and quantum fluctuation effects for frustrated systems.
To obtain the ground state of frustrated systems is to find how to put the unsatisfied bonds represented by the crosses.
Since the unsatisfied bonds are regarded as some kind of constraints,
this situation is similar with the traveling salesman problem in which there are some constraints as mentioned before.
We explain two topics in this section.
In the first half, we consider the order by disorder effect in fully-frustrated systems.
In the last half, we explain non-monotonic dynamics in decorated bond systems.

\subsubsection{Thermal Fluctuation and Quantum Fluctuation Effect of Geometrical Frustrated Systems}

\begin{figure}[t]
 \begin{center}
 \psfig{file=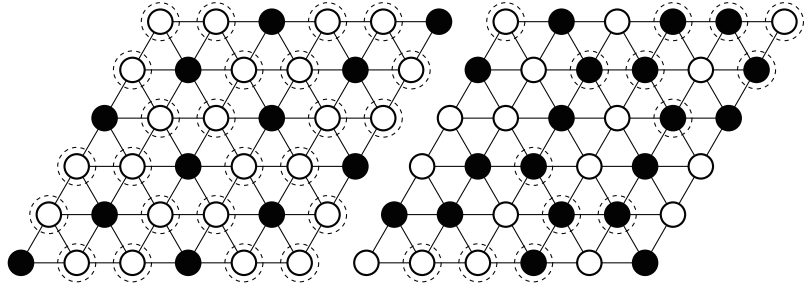, scale=0.8}
  \end{center}
 \caption{
 Typical configurations of ground states of antiferromagnetic Ising spin system on triangular lattice.
 The open and solid circles are the $+1$-state and the $-1$-state, respectively.
 The dotted circles indicate free spin where the molecular field is zero.
 }
 \label{STfig:triangularlatticegs}
\end{figure}

In general, there are many degenerated ground states in geometrical frustrated systems such as triangular antiferromagnetic Ising spin systems and kagom\'e antiferromagnetic Ising spin systems.
In these cases, non-zero residual entropy which is entropy at zero temperature exists.
Typical configurations of ground states of the triangular antiferromagnetic Ising spin systems are shown in Fig.~\ref{STfig:triangularlatticegs}.
The residual entropy per spin of this system is $S_{\rm res}^{\rm (tri)}\simeq 0.323 k_{\rm B}$\cite{STHusimi-1949,STHoutappel-1950,STWannier-1950,STWannier-1973}, where $k_{\rm B}$ is the Boltzmann constant.
Since the total entropy per spin is $k_{\rm B}\ln 2 \simeq 0.693 k_{\rm B}$, $46.6\%$ of the total entropy remains even at zero temperature.
In other words, there are macroscopic degenerated ground states in this system.
In the antiferromagnetic Ising spin system on kagom\'e lattice, there are also macroscopic degenerated ground states.
The residual entropy per spin of this system is $S_{\rm res}^{\rm (kag)}\simeq 0.502 k_{\rm B}$, which is $72.4\%$ of the total entropy\cite{STKano-1953}.

%

%
%

Suppose we apply the simulated annealing or the quantum annealing with slow schedule for geometrical frustrated spin systems.
Since there are macroscopically degenerated ground states in these systems, our purpose is to clarify whether all ground states are obtained with the same probabilities or biased probabilities.
We first consider the obtained ground states in the case of the simulated annealing with slow schedule.
If we decrease temperature slow enough, the obtained state should satisfy the equilibrium probability distribution.
When the temperature is $k_{\rm B}T \ll |J|$, the equilibrium probabilities of the ground states are dominant and that of any excited states can be neglected.
The principle of equal weight which is the keystone in the equilibrium statistical physics says that if the eigenenergies of the microscopic state $\Sigma_A$ and $\Sigma_B$ are the same, the equilibrium probability of $\Sigma_A$ and that of $\Sigma_B$ are also the same.
Then we obtain all macroscopic degenerated ground states with the same probability after the simulated annealing with slow schedule.

Next we consider the obtained ground states in the case of the quantum annealing where the transverse field decreases slow enough.
Here we assume that the initial state is set to be the ground state of the Hamiltonian at the initial time.
In order to capture the feature of the ground states in a graphical way,
it is convenient to introduce the concept of free spin where the molecular field is zero.
The molecular field at the $i$-th site is given by
\begin{eqnarray}
 h_i^{\rm (eff)} := J \sum_{j}{}^\prime \sigma_j^z,
\end{eqnarray}
where the summation runs over the nearest-neighbor sites of the $i$-th site.
For instance, in Fig.~\ref{STfig:triangularlatticegs}, spins indicated by dotted circles are free spins.
Here, the transverse field is expressed as
\begin{eqnarray}
 -\Gamma \sum_i \hat{\sigma}_i^x = 
  -\Gamma \sum_i (\hat{\sigma}_i^+ + \hat{\sigma}_i^-),
\end{eqnarray}
where $\hat{\sigma}_i^+$ and $\hat{\sigma}_i^-$ denote the raising and lowering operators at the $i$-th site, respectively.
They are defined by
\begin{eqnarray}
 \hat{\sigma}^+ := 
  \left(
   \begin{array}{cc}
    0 & 1 \\
    0 & 0 
   \end{array}
  \right),
  \qquad
  \hat{\sigma}^- :=
  \left(
   \begin{array}{cc}
    0 & 0\\
    1 & 0
   \end{array}
  \right).
\end{eqnarray}
The $x$-component of the Pauli matrix corresponds to the operator which flips the considered spin:
\begin{eqnarray}
 \hat{\sigma}^x \ket{\uparrow} = \ket{\downarrow},
  \qquad
 \hat{\sigma}^x \ket{\downarrow} = \ket{\uparrow}.
\end{eqnarray}
From this, the states which have large number of free spins are expected to become stable at the limit of $\Gamma \to 0+$ and $T=0$.
Actually, in the adiabatic limit, the amplitudes of the states which have the maximum number of free spins are larger than the others\cite{STMatsuda-2009,STMatsuda-2009b,STTanaka-2010book,STTanaka-2011c,STTanakaTamura-inprep}.
When we decrease the transverse field slow enough, the state at each time can be well approximated by the ground state of the instantaneous Hamiltonian.
Then we obtain specific ground states with high probability after the quantum annealing with slow schedule.

In this section, we considered the thermal fluctuation effect and the quantum fluctuation effect in the adiabatic limit.
The simulated annealing can obtain all the ground states with the same probability, while on the other hand, the quantum annealing can obtain specific ground states in this limit.
The biased probability distribution can be explained by the character of the quantum Hamiltonian.
The selected states should depend on how to choice the quantum Hamiltonian.
When we adopt the exchange type interaction as the quantum field, the states that have the maximum value of the ``free spin pair'' should be selected.
Moreover, it is an interesting topic to investigate differences between the simulated annealing and the quantum annealing with finite speed not only in terms of the quantum annealing but also in nonequilibrium statistical physics and condensed matter physics.
At the present stage, to consider dynamic phenomena in strongly correlated systems is difficult, since a small number of theoretical methods for obtaining dynamic phenomena have been developed.
If the technology of the artificial lattices develops more than ever, real-time dynamics and time-dependent phenomena of frustrated spin systems can be observed in real experiments.

\subsubsection{Non-Monotonic Behavior of Correlation Function in Decorated Bond System}

\begin{figure}[b]
 \begin{center}
  \psfig{file=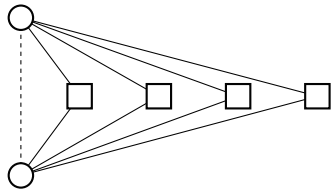, scale=0.8}
 \end{center}
 \caption{
 Decorated bond system where the number of system spins is two and the number of decorated spins is four ($N_{\rm d}=4$).
 The circles and squares represent system spins and decorated spins, respectively.
 The dotted and solid lines indicate the direct interaction between system spins and the decorated bonds, respectively.
 }
 \label{STfig:decoratedunit}
\end{figure}

In the ferromagnetic Ising spin systems, the correlation function behaves monotonic against the temperature and transverse field.
However, the behavior of the correlation function is non-monotonic as a function of temperature in some frustrated spin systems.
As an example of non-monotonic correlation function,
we introduce equilibrium properties of the correlation function in decorated bond systems in which the frustration exists.
The Hamiltonian of the decorated bond systems where the number of system spins is two shown in Fig.~\ref{STfig:decoratedunit} is given by
\begin{eqnarray}
 {\cal H} = -J_{\rm dir} \sigma_1^z \sigma_2^z - J \sum_{i=1}^{N_{\rm d}} s_i^z (\sigma_1^z + \sigma_2^z),
\end{eqnarray}
where $\sigma_i^z =\pm 1$ and $s_i^z =\pm 1$ are, respectively, called system spins and decorated spins, and $N_{\rm d}$ is the number of decorated spins.
The circles and the squares in Fig.~\ref{STfig:decoratedunit} represent the system spins and the decorated spins, respectively.

When the direct interaction between system spins $J_{\rm dir}$ is zero and the decorated bond $J$ is positive, the correlation function between system spins $\langle \sigma_1^z \sigma_2^z \rangle$ is always positive and monotonic decaying function against the temperature.
When the direct interaction between system spins $J_{\rm dir}$ is negative and the decorated bond $J$ is zero, on the other hand, the correlation function $\langle \sigma_1^z \sigma_2^z \rangle$ is always negative and monotonic increasing function against the temperature.
From this, the correlation function $\langle \sigma_1^z \sigma_2^z \rangle$ is expected to behave non-monotonic in some cases for negative $J_{\rm dir}$ and positive $J$ or positive $J_{\rm dir}$ and negative $J$.
In order to obtain temperature dependence of the correlation function between system spins, we trace over spin states except the system spins:
\begin{eqnarray}
 {\rm Tr}_{\{s_i^z\}} {\rm e}^{-\beta {\cal H}} = A {\rm e}^{K_{\rm eff}\sigma_1^z \sigma_2^z},
\end{eqnarray}
where $A$ is just a constant which does not affect any physical quantities and the effective coupling $K_{\rm eff}$ is given by
\begin{eqnarray}
 K_{\rm eff} = \frac{N_{\rm d}}{2} \ln \cosh (2\beta J) + \beta J_{\rm dir}.
\end{eqnarray}
Temperature dependence of the correlation function between system spins is represented by using $K_{\rm eff}$:
\begin{eqnarray}
 \nonumber
 C^{\rm (c)}(T) :=
  \langle \sigma_1^z \sigma_2^z \rangle 
  = \frac{
  {\rm Tr}\, \sigma_1^z \sigma_2^z {\rm e}^{-\beta{\cal H}}
  }{
  {\rm Tr}\, {\rm e}^{-\beta {\cal H}}
  }
  = \frac{
  {\rm Tr}\, \sigma_1^z \sigma_2^z {\rm e}^{K_{\rm eff}\sigma_1^z \sigma_2^z}
  }{
  {\rm Tr}\, {\rm e}^{K_{\rm eff} \sigma_1^z \sigma_2^z}
  }
  = \tanh K_{\rm eff}.\\
\end{eqnarray}
Hereafter we set $J$ as the energy unit and $J$ is positive.
In order to compare the effect of the direct interaction $J_{\rm dir}$ fairly, we assume the form such as $J_{\rm dir} = -x N_{\rm d} J$.
This is because the effective coupling $K_{\rm eff}$ is proportional to the number of decorated spins $N_{\rm d}$ under the assumption.

\begin{figure}[b]
 \begin{center}
  \psfig{file=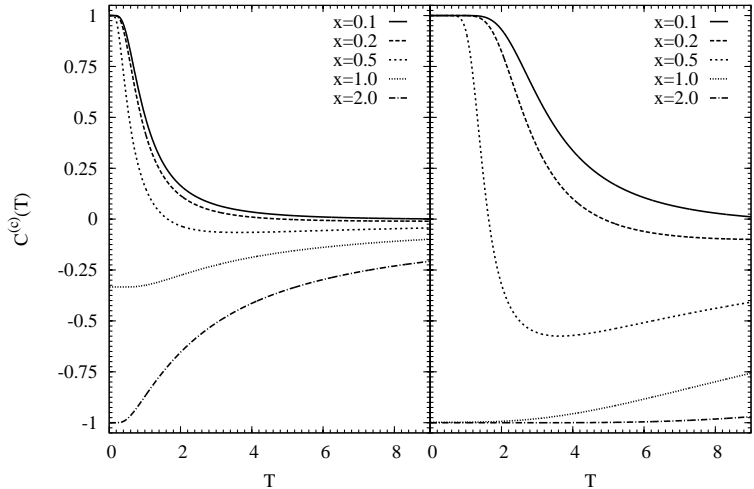, scale=0.8}
 \end{center}
 \caption{
 The correlation function between system spins $C^{\rm (c)}(T)$ as a function of temperature for $N_{\rm d}=1$ (left panel) and for $N_{\rm d}=10$ (right panel) in the cases of $x=0.1$, $0.2$, $0.5$, $1.0$, and $2.0$.
 }
 \label{STfig:correlation_T_x}
\end{figure}

Figure~\ref{STfig:correlation_T_x} shows temperature dependence of correlation function between the system spins for $N_{\rm d}=1$ and $N_{\rm d}=10$ for several $x$.
For small $x$ and large $x$, the correlation function $C^{\rm (c)}(T)$ is monotonic decreasing and increasing functions, respectively, against the temperature.
However, the correlation function $C^{\rm (c)}(T)$ behaves non-monotonic as a function of temperature for intermediate $x$.
At the temperatures where the effective coupling $K_{\rm eff}$ is larger than the critical value of the ferromagnetic Ising spin system on square lattice\cite{STOnsager-1944} $K_{\rm c}^{\rm (square)} = \frac{1}{2} \ln (1+\sqrt{2})$, ferromagnetic phase appears.
On the other hand, at the temperature where $K_{\rm eff}$ is less than $-K_{\rm c}^{\rm (square)}$, antiferromagnetic phase appears.
In this case, successive phase transitions such as paramagnetic $\to$ antiferromagnetic $\to$ paramagnetic $\to$ ferromagnetic phases occur.
Such phase transitions are called reentrant phase transitions which are sometimes appeared in frustrated systems\cite{STFradkin-1976,STMiyashita-1983,STKitatani-1985,STKitatani-1986,STAzaria-1987,STMiyashita-2001,STTanaka-2005,STTanaka-2010}.

\begin{figure}[b]
  \begin{center}
   \psfig{file=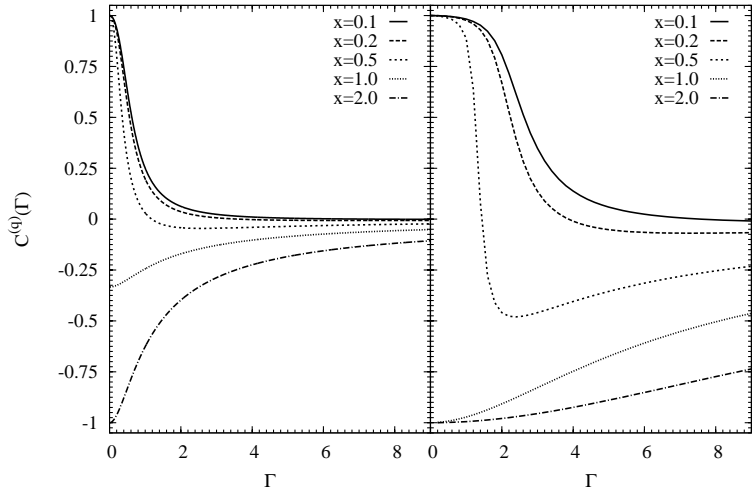, scale=0.8}
  \end{center}
 \caption{
  The correlation function between system spins $C^{\rm (q)}(\Gamma)$ as a function of transverse field for $N_{\rm d}=1$ (left panels) and for $N_{\rm d}=10$ (right panels) in the cases of $x=0.1$, $0.2$, $0.5$, $1.0$, and $2.0$.
 }
 \label{STfig:correlation_G_x}
\end{figure}

We consider transverse field response of the decorated bond systems in the ground state.
The Hamiltonian of the decorated bond system with transverse field is expressed as
\begin{eqnarray}
 \hat{{\cal H}} = 
  -J_{\rm dir} \hat{\sigma}_1^z \hat{\sigma}_2^z - J \sum_{i=1}^{N_{\rm d}} \hat{s}_i^z (\hat{\sigma}_1^z + \hat{\sigma}_2^z)
  - \Gamma (\hat{\sigma}_1^x + \hat{\sigma}_2^x + \sum_{i=1}^{N_{\rm d}} \hat{s}_i^x),
\end{eqnarray}
where $\hat{s}_i^\alpha$ denotes the $\alpha$-component of the Pauli matrix of the $i$-th decorated spin.
Here we consider transverse-field dependence of the correlation function in the ground state given by
\begin{eqnarray}
 C^{\rm (q)}(\Gamma):=\braket{\psi^{\rm (gs)}(\Gamma) | \hat{\sigma}_1^z \hat{\sigma}_2^z | \psi^{\rm (gs)}(\Gamma)},
\end{eqnarray}
where $\ket{\psi^{\rm (gs)}(\Gamma)}$ denotes the ground state at the transverse field $\Gamma$.
Figure~\ref{STfig:correlation_G_x} shows transverse-field dependence of $C^{\rm (q)}(\Gamma)$ for $N_{\rm d}=1$ and $N_{\rm d}=10$ for several $x$.

For small $x$ and large $x$, the correlation function $C^{\rm (q)}(\Gamma)$ behaves monotonic decreasing and increasing, respectively as a function of transverse field, whereas for intermediate $x$, transverse-field dependence of the correlation function behaves nonmonotonic as well as the case of thermal fluctuation.
Then, the reentrant phase transition also occurs by changing the transverse field.
However there is a difference between the thermal fluctuation effect and the quantum fluctuation effect for decorated bond system.
The temperature where $C^{\rm (c)}(T)=0$ is satisfied is the same when we change the number of decorated spins $N_{\rm d}$, whereas the transverse field at $C^{\rm (q)}(\Gamma)=0$ is different when $N_{\rm d}$ is changed.

The thermal fluctuation and the quantum fluctuation have similar properties for the phase transition phenomena in general.
Indeed, the reentrant phase transitions occur by changing the thermal fluctuation and also the quantum fluctuation as shown in this section.
However as described in Sec.~\ref{STsec:KZmechanism}, in order to obtain the best solution of optimization problems, it is better to erase phase transition.
By dealing with thermal and quantum fluctuation effects for frustrated systems exhaustively, we can construct the best form of the adding fluctuation which erases phase transition\footnote{It is not necessary that the adding fluctuation is restricted in quantum physics. From a viewpoint of optimization problems, we can arbitrary form adding term. Furthermore, it has studied that other novel fluctuation which may be able to erase phase transition as an alternative to thermal and quantum fluctuations\cite{STTamura-2010,STTanaka-2011Potts_a,STTanaka-2011a,STTanaka-2011b}.
Of course, if we want to realize experimentally, it is better that the added fluctuation term should be some kind of quantum fluctuation.}.
\section{Conclusion}

In this paper, we described some aspects of the quantum annealing from viewpoints of statistical physics, condensed matter physics, and computational physics.
Originally, the quantum annealing has been proposed as a method which can solve efficiently optimization problems in a generic way.
Since many optimization problems can be mapped onto the Ising model or generalized Ising model such as the clock model and the Potts model, it has been considered that we can obtain a better solution by using methods which were developed in computational physics.
For instance, we can obtain a better solution by decreasing temperature (thermal fluctuation) gradually in the simulated annealing which is one of the most famous practical methods.
In the quantum annealing, we decrease an introduced quantum field (quantum fluctuation) instead of temperature (thermal fluctuation).
In many studies, it was reported that a better solution can be obtained by the quantum annealing efficiently in comparison with the simulated annealing as we explained in Sec.~\ref{STsec:optimization_problems}.
Thus, the quantum annealing method is expected to be a generic and powerful solver of optimization problems as an alternative to the simulated annealing.

The quantum annealing has become a milestone of some related fields under the situation in which the quantum annealing itself has been studied exhaustively.
Since we use the quantum fluctuation in the quantum annealing with ingenuity, to obtain a better solution by using the quantum annealing is a kind of quantum information processing.
Thus, many implementation methods of the quantum annealing in theoretical and experimental ways have been proposed by many researchers.
A number of theoretical implementation methods are proposed based on knowledge of statistical physics.
As we shown in Sec.~\ref{STsec:QA_SM}, question of what are differences between the simulated annealing and the quantum annealing and question of which is efficient in the given optimization problem are catalysts to investigate differences between the thermal fluctuation and the quantum fluctuation in a deeper way.
On top of that, studies on the quantum annealing are expected to open the door to consider equilibrium and nonequilibrium statistical physics.
Recently, preparation methods of intended Hamiltonian have been established in some experimental systems such as artificial lattices and nuclear magnetic resonance because of recent development of experimental techniques.
As long as we use classical computer and our present knowledge, there are a huge number of problems where to obtain the best solution is difficult without any and every approximation in theoretical methods.
However if we prepare the Hamiltonian which expresses our intended problem, we can {\it calculate} experimentally the stable state of the prepared Hamiltonian in near future.

The quantum annealing transcends just a method for obtaining the best solution of optimization problems and it will make a development in wide area of science.
Although it seems that studies on the quantum annealing itself have been well established, we believe that the quantum annealing plays a role as a bridge with the abovementioned area of science and the quantum information.

\section*{Acknowledgement}

The authors are grateful to Bernard Barbara, Bikas K. Charkrabarti, Naomichi Hatano, Masaki Hirano, Naoki Kawashima, Kenichi Kurihara, Yoshiki Matsuda, Seiji Miyashita, Hiroshi Nakagawa, Mikio Nakahara, Hidetoshi Nishimori, Masayuki Ohzeki, Hans de Raedt, Per Arne Rikvold, Issei Sato, Sei Suzuki, Eric Vincent, and Yoshihisa Yamamoto for their valuable comments.
S.T. acknowledges Keisuke Fujii, Yoshifumi Nakada, and Takahiro Sagawa for their useful discussion during the lecture.
S.T. is partly supported by Grand-in-Aid for JSPS Fellows (23-7601).
R.T. is partly supported financially by National Institute for Materials Science (NIMS).
The computation in the present work was performed on computers at the Suprecomputer Center, Institute for Solid State Physics, University of Tokyo.

\bibliographystyle{ws-procs9x6}
\bibliography{ws-pro-sample}

\begin{thebibliography}{999}

%
%

\bibitem{STKirkpatrick-1983}
	S.~Kirkpatrick, C.~D.~Gelatt Jr., and M.~P.~Vecchi,
	{\em Science} {\bf 220}, 671 (1983).

\bibitem{STKirkpatrick-1984}
	S.~Kirkpatrick,
	{\em J. Stat. Phys.} {\bf 34}, 975 (1984).

\bibitem{STGeman-1984}
	S.~Geman and D.~Geman,
	{\em IEEE Transactions on Pattern Analysis and Machine Intelligence} {\bf 6}, 721 (1984).

\bibitem{STFinnila-1994}
	A.~B.~Finnila, M.~A.~Gomez, C.~Sebenik, C.~Stenson, and J.~D.~Doll,
	{\em Chem. Phys. Lett.} {\bf 219}, 343 (1994).

\bibitem{STKadowaki-1998a}
	T.~Kadowaki and H.~Nishimori,
	{\em Phys. Rev. E} {\bf 58}, 5355 (1998).

\bibitem{STBrooke-1999}
	J.~Brooke, D.~Bitko, T.~F.~Rosenbaum, and G.~Aeppli,
	{\em Science} {\bf 284}, 779 (1999).

\bibitem{STFarhi-2001}
	E.~Farhi, J.~Goldstone, S.~Gutmann, J.~Lapan, A.~Lundgren, and D.~Preda,
	{\em Science} {\bf 292}, 472 (2001).

\bibitem{STSantoro-2002}
	G.~E.~Santoro, R.~Marto\v{n}\'ak, E.~Tosatti, and R.~Car,
	{\em Science} {\bf 295}, 2427 (2002). 

\bibitem{STDas-2005}
	A.~Das and B.~K.~Chakrabarti,
	{\it Quantum Annealing and Related Optimization Methods} (Springer, Heidelberg, 2005).

\bibitem{STDas-2008}
	A.~Das and B.~K.~Chakrabarti,
	{\em Rev. Mod. Phys.} {\bf 80}, 1061 (2008).

\bibitem{STOhzeki-2011}
	M.~Ohzeki and H.~Nishimori,
	{\em J. Comp. and Theor. Nanoscience} {\bf 8}, 963 (2011).

\bibitem{STKurihara-2009}
	K.~Kurihara, S.~Tanaka, and S.~Miyashita,
	{\em Proceedings of the 25th Conference on Uncertainty in Artificial Intelligence} (2009).

\bibitem{STSato-2009}
	I.~Sato, K.~Kurihara, S.~Tanaka, H.~Nakagawa, and S.~Miyashita,
	{\em Proceedings of the 25th Conference on Uncertainty in Artificial Intelligence} (2009).

\bibitem{STTanaka-2011a}
	S.~Tanaka, R.~Tamura, I.~Sato, and K.~Kurihara,
	to appear in {\em Kinki University Quantum Computing Series: ``Summer School on Diversities in Quantum Computation/Information''}.

\bibitem{STJarzynski-1997a}
	C.~Jarzynski,
	{\em Phys. Rev. Lett.} {\bf 78}, 2690 (1997).

\bibitem{STJarzynski-1997b}
	C.~Jarzynski,
	{\em Phys. Rev. E} {\bf 56}, 5018 (1997).

\bibitem{STOhzeki-2010}
	M.~Ohzeki,
	{\em Phys. Rev. Lett.} {\bf 105}, 050401 (2010).

%
%

%
%

\bibitem{STIsing-1925}
	E.~Ising, 
	{\em Z. Phys.} {\bf 31}, 253 (1925).

\bibitem{STOnsager-1944}
	L.~Onsager,
	{\em Phys. Rev.} {\bf 65}, 117 (1944).

\bibitem{STBlume-1966}
	M.~Blume,
	{\em Phys. Rev.} {\bf 141}, 517 (1966).

\bibitem{STCapel-1966}
	H.~W.~Capel,
	{\em Phys. Lett.} {\bf 23}, 327 (1966).

\bibitem{STTobochnik-1982}
	J.~Tobochnik,
	{\em Phys. Rev. B} {\bf 26}, 6201 (1982).

\bibitem{STChalla-1986}
	M.~S.~S.~Challa and D.~P.~Landau,
	{\em Phys. Rev. B} {\bf 33}, 437 (1986).

\bibitem{STPotts-1952}
	R.~B.~Potts,
	{\em Proc. Cambridge Philos. Soc.} {\bf 48}, 106 (1952).

\bibitem{STWu-1982}
	F.~Y.~Wu,
	{\em Rev. Mod. Phys.} {\bf 54}, 235 (1982).

\bibitem{STOhtsuka-1961}
T. Ohtsuka, {\it J. Phys. Soc. Jpn.} \textbf{16}, 1549 (1961). 
\bibitem{STRayl-1968}
M. Rayl, O. E. Vilches, and J. C. Wheatley, {\it Phys. Rev.} \textbf{165}, 698 (1968).
\bibitem{STOno-1970}
K. \^{O}no, M. Shinohara, A. Ito, N. Sakai, and M. Suenaga, {\it Phys. Rev. Lett.} \textbf{24}, 770 (1970).

\bibitem{STAchiwa-1969}
N. Achiwa, {\it J. Phys. Soc. Jpn.} \textbf{27}, 561 (1969).
\bibitem{STMekata-1978}
M. Mekata and K. Adachi, {\it J. Phys. Soc. Jpn.} \textbf{44}, 806 (1978).

\bibitem{STCooke-1959}
A. H. Cooke, D. T. Edmonds, F. R. McKim, and W. P. Wolf, {\it Proc. Roy. Soc. London Ser. A} \textbf{252}, 246 (1959).
\bibitem{STCooke-1968a}
A. H. Cooke, D. T. Edmonds, C. B. P. Finn, and W. P. Wolf, {\it Proc. Roy. Soc. London Ser. A}  \textbf{306}, 313 (1968).
\bibitem{STCooke-1968b}
A. H. Cooke, D. T. Edmonds, C. B. P. Finn, and W. P. Wolf, {\it Proc. Roy. Soc. London Ser. A}  \textbf{306}, 335 (1968). 

\bibitem{STTakeda-1970}
K. Takeda, M. Matsuura, S. Matsukawa, Y. Ajiro, and T. Haseda, {\it Proc. 12th Int. Conf. Low Temp. Phys., Kyoto} 803 (1970).
\bibitem{STTakeda-1971a}
K. Takeda, S. Matsukawa, and T. Haseda, {\it J. Phys. Soc. Jpn.} \textbf{30}, 1330 (1971).


\bibitem{STFiggis-1964}
B. N. Figgis, M. Gerloch, and R. Mason, {\it Acta. Crystallogr.} \textbf{17}, 506 (1964).
\bibitem{STWielinga-1967}
R. F. Wielinga, H. W. J. Blote, J. A. Roest, and W. J. Huiskamp, {\it Physica}  \textbf{34}, 223 (1967).
\bibitem{STMess-1967}
K. W. Mess, E. Lagendijk, D. A. Curtis, and W. J. Huiskamp, {\it Physica} \textbf{34}, 126 (1967).

\bibitem{STHoy-1965}
G. R. Hoy and F. de S. Barros, {\it Phys. Rev.} \textbf{139}, A929 (1965).
\bibitem{STMatsuura-1970}
M. Matsuura, H. W. J. Blote, and W. J. Huiskamp, {\it Physica} \textbf{50}, 444 (1970).
\bibitem{STPierce-1971}
R. D. Pierce and S. A. Friedberg, {\it Phys. Rev. B} \textbf{3}, 934 (1971).
\bibitem{STTakeda-1971b}
K. Takeda and S. Matsukawa, {\it J. Phys. Soc. Jpn.} \textbf{30}, 887 (1971).

\bibitem{STStryiewski-1977}
E. Stryjewski and N. Giordano, {\it Adv. Phys.} \textbf{26}, 487 (1977).
\bibitem{STBreed-1969}
D. J. Breed, K. Gilijamse, and A. R. Miedema, {\it Physica} \textbf{45}, 205 (1969).

\bibitem{STOno-1964}
K. \^Ono, A. Ito, and T. Fujita, {\it J. Phys. Soc. Jpn.} \textbf{19}, 2119 (1964).
\bibitem{STBirgeneau-1972}
R. J. Birgeneau, W. B. Yelon, E. Cohen, and J. Makovsky, {\it Phys. Rev. B} \textbf{5}, 2607 (1972).



\bibitem{STWright-1971}
J. C. Wright, H. W. Moos, J. H. Colwell, B. W. Magnum, and D. D. Thornton, {\it Phys. Rev. B} \textbf{3}, 843 (1971).
\bibitem{STRado-1969}
G. T. Rado, {\it Phys. Rev. Lett.} \textbf{23}, 644 (1969).
\bibitem{STScharenberg-1971}
W. Scharenberg and G. Will, {\it Int. J. Magnetism} \textbf{1}, 277 (1971).
\bibitem{STFuess-1971}
H. Fuess, A. Kallel, and F. Tch\'eou, {\it Solid State Commun.} \textbf{9}, 1949 (1971).

\bibitem{STBall-1963}
M. Ball, M. J. M. Leask, W. P. Wolf, and A. F. G. Wyatt, {\it J. Appl. Phys.} \textbf{34}, 1104 (1963).
\bibitem{STNorvell-1969a}
J. C. Norvell, W. P. Wolf, L. M. Corliss, J. M. Hastings, and R. Nathans, {\it Phys. Rev.} \textbf{186}, 557 (1969). 
\bibitem{STNorvell-1969b}
J. C. Norvell, W. P. Wolf, L. M. Corliss, J. M. Hastings, and R. Nathans, {\it Phys. Rev.} \textbf{186}, 567 (1969). 

\bibitem{STBaker-1963}
G. A. Baker, Jr., {\it Phys. Rev.} \textbf{129}, 99 (1963).
\bibitem{STSykes-1972}
M. F. Sykes, D. L. Hunter, D. S. McKenzie, and B. R. Heap, {\it J. Phys. A: Gen. Phys.} \textbf{5}, 667 (1972).

\bibitem{STStout-1955}
J. W. Stout and E. Catalano, {\it J. Chem. Phys.} \textbf{23}, 2013 (1955).
\bibitem{STDomb-1964}
C. Domb and A. R. Miedema, {\it Progress in low Temperature Physics, Vol. 4, edited by C. J. Gorter} (North-Holland, Amsterdam, 1964).
\bibitem{STWertheim-1967}
G. K. Wertheim and D. N. E. Buchanan, {\it Phys. Rev.} \textbf{161}, 478 (1967).
\bibitem{STShapira-1970}
Y. Shapira, {\it Phys. Rev. B} \textbf{2}, 2725 (1970).

%
\bibitem{STNielsen-2000}
M.~A.~Nielsen and I.~L.~Chuang,
{\it Quantum Computation and Quantum Information}
(Cambridge University Press, Cambridge, 2000).
%
\bibitem{STNakahara-2008}
M.~Nakahara and T.~Ohmi,
{\it Quantum Computing: From Linear Algebra to Physical Realizations}
(Taylor \& Francis, London, 2008).
%
\bibitem{STCory-1997}
D.~G.~Cory, A.~F.~Fahmy, and T.~F.~Havel,
{\em Proc. Natl. Acad. Sci. USA} {\bf 94}, 1634 (1997).
%
\bibitem{STCory-1998}
D.~G.~Cory, M.~D.~Price, W.~Maas, E.~Knill, R.~Laflamme, W.~H.~Zurek, T.~F.~Havel, and S.~S.~Somaroo,
{\em Phys. Rev. Lett.} {\bf 81}, 2152 (1998).
%
\bibitem{STGershenfeld-1997}
N.~A.~Gershenfeld and I.~L.~Chuang,
{\em Science} {\bf 275}, 350 (1997).
%
\bibitem{STChuang-1998}
I.~L.~Chuang, L.~M.~K.~Vandersypen, X.~Zhou, D.~W.~Leung, and S.~Lloyd,
{\em Nature} {\bf 393}, 143 (1998).
%
\bibitem{STJones-1998}
J.~A.~Jones and M.~Mosca,
{\em J. Chem. Phys.} {\bf 109}, 1648 (1998).
%
\bibitem{STKnill-1998}
E.~Knill, I.~Chuang, and R.~Laflamme,
{\em Phys. Rev. A} {\bf 57}, 3348 (1998).
%
\bibitem{STLaflamme-1998}
R.~Laflamme, E.~Knill, W.~H.~Zurek, P.~Catasti, and S.~V.~S.~Mariappan,
{\em Phil. Trans. R. Soc. Lond. A} {\bf 356}, 1941 (1998).
%
\bibitem{STJones-1999}
J.~A.~Jones and M.~Mosca,
{\em Phys. Rev. Lett.} {\bf 83}, 1050 (1999).
%
\bibitem{STPrice-1999}
M.~D.~Price, S.~S.~Somaroo, A.~E.~Dunlop, T.~F.~Havel, and D.~G.~Cory,
{\em Phys. Rev. A} {\bf 60}, 2777 (1999).
%
\bibitem{STVandersypen-1999}
L.~M.~K.~Vandersypen, C.~S.~Yannoni, M.~H.~Sherwood, and I.~L.~Chuang,
{\em Phys. Rev. Lett.} {\bf 83}, 3085 (1999).
%
\bibitem{STVandersypen-2000}
L.~M.~K.~Vandersypen, M.~Steffen, G.~Breyta, C.~S.~Yannoni, R.~Cleve, and I.~L.~Chuang,
{\em Phys. Rev. Lett.} {\bf 85}, 5452 (2000).
%
\bibitem{STVandersypen-2001}
L.~M.~K.~Vandersypen, M.~Steffen, G.~Breyta, C.~S.~Yannoni, M.~H.~Sherwood, and I.~L.~Chuang,
{\em Nature} (London) {\bf 414}, 883 (2001).
%
\bibitem{STNakahara-2004}
M.~Nakahara, Y.~Kondo, K.~Hata, and S.~Tanimura,
{\em Phys. Rev. A} {\bf 70}, 052319 (2004).
%
\bibitem{STKondo-2007}
Y.~Kondo,
{\em J. Phys. Soc. Jpn.} {\bf 76}, 104004 (2007).
%

%
%

%
%

\bibitem{STSuwa-2010}
	H.~Suwa and S.~Todo,
	{\em Phys. Rev. Lett.} {\bf 105}, 120603 (2010).

\bibitem{STSuwa-2011}
	H.~Suwa and S.~Todo,
	{\em arXiv}:1106.3562.

\bibitem{STSwendsen-1987}
	R.~H.~Swendsen and J.~S.~Wang,
	{\em Phys. Rev. Lett.} {\bf 58}, 86 (1987).

\bibitem{STWolff-1989}
	U.~Wolff,
	{\em Phys. Rev. Lett.} {\bf 62}, 361 (1989).

\bibitem{STHukushima-1996}
	K.~Hukushima and K.~Nemoto,
	{\em J. Phys. Soc. Jpn.} {\bf 65}, 1604 (1996).

\bibitem{STHarada-2004}
	N.~Kawashima and K.~Harada,
	{\em J. Phys. Soc. Jpn.} {\bf 73}, 1379 (2004).

\bibitem{STNakamura-2008}
	T.~Nakamura,
	{\em Phys. Rev. Lett.} {\bf 101}, 210602 (2008).

\bibitem{STMorita-2009}
	S.~Morita, S.~Suzuki, and T.~Nakamura,
	{\it Phys. Rev. E} {\bf 79}, 065701(R) (2009).


\bibitem{STTrotter-1959}
	H.~F.~Trotter,
	{\em Proc. Am. Math. Soc.} {\bf 10}, 545 (1959).

\bibitem{STSuzuki-1976}
	M.~Suzuki,
	{\em Prog. Theor. Phys.} {\bf 56}, 1454 (1976).


\bibitem{STKadowaki-1998b}
	T.~Kadowaki,
	{\em Ph. D thesis, Tokyo Institute of Technology} (1998).



\bibitem{STTanaka-2000}
	K.~Tanaka and T.~Horiguchi,
	{\em Electronics and Communications in Japan, Part 3: Fundamental Electronic Science} {\bf 83}, 84 (2000).

\bibitem{STTanaka-2002}
	K.~Tanaka and T.~Horiguchi,
	{\em Interdisciplinary Information Science} {\bf 8}, 33 (2002).

\bibitem{STAttias-1999}
	H.~Attias,
	{\em Proceedings of the 15th Conference on Uncertainly in Artificial Intelligence} 21 (1999).


\bibitem{STLandau-1932}
	L.~Landau,
	{\em Phys. Z. Sowjetunion} {\bf 2}, 46 (1932).

\bibitem{STZener-1932}
	C.~Zener,
	{\em Proc. R. Soc. London Ser. A} {\bf 137}, 696 (1932).

\bibitem{STStuckelberg-1932}
	E.~C.~G.~St\"uckelberg,
	{\em Helv. Phys. Acta} {\bf 5}, 369 (1932).

\bibitem{STRosen-1932b}
	N.~Rosen and C.~Zener,
	{\em Phys. Rev.} {\bf 40}, 502 (1932).


\bibitem{STChakrabarti-1996}
B.~K.~Chakrabarti, A.~Dutta, and P.~Sen,
{\it Quantum Ising Phases and Transitions in Transverse Ising Models}
(Springer Verlag, Berlin, 1996).
%
\bibitem{STTrammel-1960}
G.~T.~Trammel,
{\em J. Appl. Phys.} {\bf 31}, 362S (1960).
%
\bibitem{STCooke-1962}
A.~H.~Cooke, D.~T.~Edmonds, C.~B.~P.~Finn, and W.~P.~Wolf,
{\em J. Phys. Soc. Jpn.} {\bf 17}, Suppl. B1 481 (1962).
%
\bibitem{STStout-1962}
J.~W.~Stout and R.~C.~Chisolm,
{\em J. Chem. Phys.} {\bf 36}, 979 (1962).
%
\bibitem{STMoruzzi-1963}
V.~L.~Moruzzi and D.~T.~Teaney,
{\em Sol. State. Comm.} {\bf 1}, 127 (1963).
%
\bibitem{STNarath-1966}
A.~Narath and J.~E.~Schriber,
{\em J. Appl. Phys.} {\bf 37}, 1124 (1966).
%
\bibitem{STWielinga-1969}
R.~F.~Wielinga and W.~J.~Huiskamp,
{\em Physica} {\bf 40}, 602 (1969).
%
\bibitem{STWolf-1971}
W.~P.~Wolf,
{\em J. Phys.} (Paris) {\em 32 Suppl.} {\bf C1} 26 (1971).
%
\bibitem{STWu-1991}
W.~Wu, B.~Ellman, T.~F.~Rosenbaum, G.~Aeppli, and D.~H.~Reich,
{\em Phys. Rev. Lett.} {\bf 67}, 2076 (1991).
%
\bibitem{STWu-1993}
W.~Wu, D.~Bitko, T.~F.~Rosenbaum, and G.~Aeppli,
{\em Phys. Rev. Lett.} {\bf 71}, 1919 (1993).
%
%
\bibitem{STReich-1990}
D.~H.~Reich, B.~Ellman, J.~Yang, T.~F.~Rosenbaum, G.~Aeppli, and D.~P.~Belanger,
{\em Phys. Rev. B} {\bf 42}, 4631 (1990).
%
\bibitem{STRosenbaum-1996}
T.~F.~Rosenbaum,
{\em J. Phys.: Condens. Matter} {\bf 8}, 9759 (1996).
%
\bibitem{STReich-1986}
D.~H.~Reich, T.~F.~Rosenbaum, G.~Aeppli, and H.~Guggenheim,
{\em Phys. Rev. B} {\bf 34}, 4956 (1986).
%
\bibitem{STMydosh-1993}
J.~A.~Mydosh,
{\it Spin Glasses: An Experimental Introduction} (Taylor \& Francis, London, 1993).
%
%
\bibitem{STBak-1987}
P.~Bak, C.~Tang, and K.~Wiesenfeld,
{\em Phys. Rev. Lett.} {\bf 59}, 381 (1987).
%




%
%

%
%

\bibitem{STMartonak-2004}
	R.~Marto\v{n}\'ak, G.~E.~Santoro, and E.~Tosatti,
	{\em Phys. Rev. E} {\bf 70}, 057701 (2004).

%
%

%
%


\bibitem{STTanaka-2007}
	S.~Tanaka and S.~Miyashita,
	{\em J. Phys.: Condens. Matter} {\bf 19}, 145256 (2007).

\bibitem{STTakayama-2007}
	H.~Takayama and K.~Hukushima,
	{\em J. Phys. Soc. Jpn.} {\bf 76}, 013702 (2007).

\bibitem{STTanaka-2007b}
	S.~Tanaka and S.~Miyashita,
	{\em J. Phys. Soc. Jpn.} {\bf 76}, 103001 (2007).

\bibitem{STMiyashita-2007}
	S.~Miyashita, S.~Tanaka, and M.~Hirano,
	{\em J. Phys. Soc. Jpn.} {\bf 76}, 083001 (2007).

\bibitem{STTanaka-2009}
	S.~Tanaka and S.~Miyashita,
	{\em J. Phys. Soc. Jpn.} {\bf 78}, 084002 (2009).

\bibitem{STTanaka-2010}
	S.~Tanaka and S.~Miyashita,
	{\em Phys. Rev. E} {\bf 81}, 051138 (2010),
	{\em Virtual Journal of Quantum Information} {\bf 10}, (2010).

\bibitem{STTanaka-2011b}
	S.~Tanaka and R.~Tamura,
	{\em J. Phys.: Conf. Ser.} {\bf 320}, 012025 (2011).

\bibitem{STNishimori-2011}
H. Nishimori and G. Ortiz,
{\it Elements of Phase Transitions and Critical Phenomena}
(Oxford Univ Press, Oxford, 2010).

\bibitem{STIto-1987}
    N.~Ito, M.~Taiji, and M.~Suzuki,
    {\em J. Phys. Soc. Jpn.} {\bf 56}, 4218 (1987).

\bibitem{STKibble-1976}
T. W. B. Kibble,
{\em J. Phys. A} \textbf{9}, 1387 (1976).

\bibitem{STKibble-1980}
T. W. B. Kibble,
{\em Phys. Rep.} \textbf{67}, 183 (1980).

\bibitem{STZurek-1985}
W. H. Zurek,
{\em Nature} (London) \textbf{317}, 505 (1985).

\bibitem{STDamski-2005}
B. Damski,
{\em Phys. Rev. Lett.} \textbf{95}, 035701 (2005).

\bibitem{STZurek-2005} 
W. H. Zurek, U. Dorner, and P. Zoller,
{\em Phys. Rev. Lett.} \textbf{95}, 105701 (2005). 

\bibitem{STRuutu-1996}
V. M. H. Ruutu, V. B. Eltsov, A. J. Gill, T. W. B. Kibble, M. Krusius, Y. G. Makhlin, B. Placais, G. E. Volovik, and W. Xu,
{\em Nature} (London) \textbf{382}, 334 (1996).

\bibitem{STEltsov-2000}
V. B. Eltsov, T. W. B. Kibble, M. Krusius, V. M. H. Ruutu, and G. E. Volovik,
{\em Phys. Rev. Lett.} \textbf{85}, 4739 (2000).  

\bibitem{STSaito-2007}
H. Saito, Y. Kawaguchi, and M. Ueda,
{\em Phys. Rev. A} \textbf{76}, 043613 (2007).

\bibitem{STWeiler-2008}
C. N. Weiler, T. W. Neely, D. R. Scherer, A. S. Bradley, M. J. Davis, and B. P. Anderson,
{\em Nature} (London) \textbf{455}, 948 (2008).

\bibitem{STSuzuki-2009}
S. Suzuki,
{\em J. Stat. Mech.} P03032 (2009).

\bibitem{STSuzuki-2011a}
S. Suzuki,
{\em J. Phys.: Conf. Ser.} \textbf{302}, 012046 (2011).

\bibitem{STGlauber-1963}
R. J. Glauber,
{\em J. Math. Phys.} \textbf{4}, 294 (1963).

\bibitem{STShankar-1987}
R. Shankar and G. Murthy,
{\em Phys. Rev. B} \textbf{36}, 536 (1987).

\bibitem{STFisher-1995}
D. S. Fisher,
{\em Phys. Rev. B} \textbf{51}, 6411 (1995).

\bibitem{STDziarmaga-2006}
J. Dziarmaga,
{\em Phys. Rev. B} \textbf{74}, 064416 (2006).


\bibitem{STBiroli-2010}
	G.~Biroli, L.~F.~Cugliandolo, and A.~Sicilia,
	{\em Phys. Rev. E} {\bf 81}, 050101(R) (2010).


\bibitem{STToulouse-1977}
	G.~Toulouse,
	{\em Commun. Phys.} (London) {\bf 2}, 115 (1977).

\bibitem{STLiebmann-1986}
	R.~Liebmann,
	{\it Statistical Mechanics of Periodic Frustrated Ising Systems} (Springer-Verlag, Berlin/Heidelberg, GmbH, Heidelberg, 1986).

\bibitem{STKawamura-1998}
	H.~Kawamura,
	{\em J. Phys.: Condens. Matter} {\bf 10}, 4707 (1998).

\bibitem{STDiep-2005}
	H.~T.~Diep (ed.),
	{\it Frustrated Spin Systems} (World Scientific, Singapore, 2005).

\bibitem{STTanaka-2005}
	S.~Tanaka and S.~Miyashita,
	{\em Prog. Theor. Phys. Suppl.} {\bf 157}, 34 (2005).

\bibitem{STTanaka-2007c}
	S.~Tanaka and S.~Miyashita,
	{\em J. Phys. Soc. Jpn.} {\bf 76}, 103001 (2007).

\bibitem{STTamura-2008}
	R.~Tamura and N.~Kawashima,
	{\em J. Phys. Soc. Jpn.} {\bf 77}, 103002 (2008).

\bibitem{STTanaka-2009b}
	S.~Tanaka and S.~Miyashita,
	{\em J. Phys. Soc. Jpn.} {\bf 78}, 084002 (2009).

\bibitem{STTamura-2011}
	R.~Tamura and N.~Kawashima,
	{\em J. Phys. Soc. Jpn.} {\bf 80}, 074008 (2011).

\bibitem{STTamura-2011b}
	R.~Tamura, N.~Kawashima, T.~Yamamoto, C.~Tassel, and H.~Kageyama,
	{\em Phys. Rev. B} {\bf 84}, 214408 (2011).


\bibitem{STHusimi-1949}
	K.~Husimi and I.~Syozi,
	{\em Prog. Theor. Phys.} {\bf 5}, 177 (1950).

\bibitem{STHoutappel-1950}
	R.~M.~F.~Houtappel,
	{\em Physica} {\bf 16}, 425 (1950).

\bibitem{STWannier-1950}
	G.~H.~Wannier,
	{\em Phys. Rev.} {\bf 79}, 357 (1950).

\bibitem{STWannier-1973}
	G.~H.~Wannier,
	{\em Phys. Rev. B} {\bf 7}, 5017 (1973).

\bibitem{STKano-1953}
	K.~Kano and S.~Naya,
	{\em Prog. Theor. Phys.} {\bf 10}, 158 (1953).

\bibitem{STMatsuda-2009}
	Y.~Matsuda, H.~Nishimori, and H.~G.~Katzgraber,
	{\em J. Phys.: Conf. Ser.} {\bf 143}, 012003 (2009).

\bibitem{STMatsuda-2009b}
	Y.~Matsuda, H.~Nishimori, and H.~G.~Katzgraber,
	{\em New J. Phys.} {\bf 11}, 073021 (2009).

\bibitem{STTanaka-2010book}
	S.~Tanaka, M.~Hirano, and S.~Miyashita,
	{\it Lecture Note in Physics ``Quantum Quenching, Annealing, and Computation''} (Springer) {\bf 802}, 215 (2010).
	
\bibitem{STTanaka-2011c}
	S.~Tanaka,
	to appear in {\it proceedings of Kinki University Quantum Computing Series: ``Symposium on Quantum Information and Quantum Computing''} (2011).

\bibitem{STTanakaTamura-inprep}
	S.~Tanaka and R.~Tamura,
	{\em in preparation}.

\bibitem{STFradkin-1976}
	E.~H.~Fradkin and T.~P.~Eggarter, 
	{\em Phys. Rev. A} {\bf 14}, 495 (1976).

\bibitem{STMiyashita-1983}
	S.~Miyashita,
	{\em Prog. Theor. Phys.} {\bf 69}, 714 (1983).

\bibitem{STKitatani-1985}
	H.~Kitatani, S.~Miyashita, and M.~Suzuki,
	{\em Phys. Lett.} {\bf 108A}, 45 (1985).

\bibitem{STKitatani-1986}
	H.~Kitatani, S.~Miyashita, and M.~Suzuki,
	{\em J. Phys. Soc. Jpn.} {\bf 55}, 865 (1986).

\bibitem{STAzaria-1987}
	P.~Azaria, H.~T.~Diep, and H.~Giacomini,
	{\em Phys. Rev. Lett.} {\bf 59}, 1629 (1987).

\bibitem{STMiyashita-2001}
	S.~Miyashita and E.~Vincent,
	{\em Eur. Phys. J. B} {\bf 22}, 203 (2001).

\bibitem{STTamura-2010}
	R.~Tamura, S.~Tanaka, and N.~Kawashima,
	{\em Prog. Theor. Phys.} {\bf 124}, 381 (2010).

\bibitem{STTanaka-2011Potts_a}
	S.~Tanaka and R.~Tamura, and N.~Kawashima,
	{\em J. Phys.: Conf. Ser.} {\bf 297}, 012022 (2011).


%
%

\end{thebibliography}

\end{document}